\documentclass[floatfix,nofootinbib,twocolumn,preprintnumbers,superscriptaddress,notitlepage,nolongbibliography,aps,prd]{revtex4-2}

\usepackage{times}
\usepackage{xcolor}
\usepackage{soul}
\usepackage{listings}
\usepackage{amsfonts}
\usepackage{amsmath}
\usepackage[justification=RaggedRight]{caption} 
\usepackage{orcidlink}
\usepackage{multirow}
\usepackage{amssymb}
\usepackage{bm}
\usepackage{dcolumn}
\usepackage{enumerate}
\usepackage{epsfig}
\usepackage{graphicx}
\usepackage{graphics}
\usepackage[latin1]{inputenc}
\usepackage{latexsym}
\usepackage{rotating}
\usepackage{hyperref}
\usepackage{cleveref}
\usepackage[caption=false]{subfig}
\usepackage[normalem]{ulem}
\usepackage[T1]{fontenc}
\usepackage{upquote}
\usepackage{etoolbox} 

\definecolor{bg}{rgb}{0.95,0.95,0.95}
\definecolor{keyword}{rgb}{0,0,0}
\definecolor{string}{rgb}{0.8,0,0}
\definecolor{comment}{rgb}{0,0.5,0}

\robustify{\texttt}
\let\originaltexttt\texttt

\begingroup
\catcode`'=\active
\catcode``=\active
\makeatletter
\renewrobustcmd{\texttt}[1]{%
   {%
   \everyeof{\noexpand}\endlinechar-1
   \expandafter\catcode\string``=\active
   \expandafter\catcode\string`'=\active
   \let'\textquotesingle
   \let`\textasciigrave
   \ifx\encodingdefault\upquote@OTone
    \ifx\ttdefault\upquote@cmtt
     \def'{\char13 }\def`{\char18 }%
    \fi
   \fi
   \scantokens{\originaltexttt{#1}}%
   }%
}%
\endgroup

\newcommand{\<}{\begin{equation}}
\newcommand{\?}{\end{equation}}

\newcommand{\cE}{\mathcal{E}}
\newcommand{\cH}{\mathcal{H}}

\newcommand{\cK}{\mathcal{K}}
\newcommand{\cL}{\mathcal{L}}

\newcommand{\cO}{\mathcal{O}}
\newcommand{\cP}{\mathcal{P}}

\newcommand{\cX}{\mathcal{X}}

\newcommand{\BS}{\mathbf{S}}
\newcommand{\Bs}{\mathbf{s}}
\newcommand{\BhL}{\mathbf{\hat{L}}}

\newcommand{\bx}{\bar{x}}

\definecolor{mgreen}{rgb}{0.1,0.7,0.1}

\begin{document}

\title{Evolution of precessing binary black holes on eccentric orbits using orbit-averaged evolution equations}
\author{Khun Sang Phukon \orcidlink{0000-0003-1561-0760}}
\email{k.s.phukon@bham.ac.uk}
\affiliation{School of Physics and Astronomy and Institute for Gravitational Wave Astronomy,\\University of Birmingham, Edgbaston, Birmingham, B15 2TT, United Kingdom}
\author{Nathan~K.~Johnson-McDaniel \orcidlink{0000-0001-5357-9480}}
\email{nkjm.physics@gmail.com}
\affiliation{Department of Physics and Astronomy, The University of Mississippi, University, Mississippi 38677, USA}
\author{Amitesh Singh \orcidlink{0000-0002-4697-1254}}
\email{amiteshsingh487@gmail.com}
\affiliation{Department of Physics and Astronomy, The University of Mississippi, University, Mississippi 38677, USA}
\author{Anuradha Gupta \orcidlink{0000-0002-5441-9013}}
\email{agupta1@olemiss.edu}
\affiliation{Department of Physics and Astronomy, The University of Mississippi, University, Mississippi 38677, USA}

\begin{abstract}
The most general bound binary black hole system has an eccentric orbit and precessing spins. The detection of such a system with significant eccentricity close to the merger would be a clear signature of dynamical formation. In order to study such systems, it is important to be able to evolve their spins and eccentricity from the larger separations at which the binary formed to the smaller separations at which it is detected, or vice versa. Knowledge of the precessional evolution of the binary's orbital angular momentum can also be used to twist up aligned-spin eccentric waveform models to create a spin-precessing eccentric waveform model. In this paper, we present a new publicly available code to evolve eccentric, precessing binary black holes using orbit-averaged post-Newtonian (PN) equations from the literature. The spin-precession dynamics is 2PN accurate, i.e., with the leading spin-orbit and spin-spin corrections. The evolution of orbital parameters (orbital frequency, eccentricity, and periastron precession), which follow the quasi-Keplerian parametrization, is 3PN accurate in the point particle terms and includes the leading order spin-orbit and spin-spin effects. All the spin-spin terms include the quadrupole-monopole interaction. The eccentricity enhancement functions in the fluxes use the high-accuracy hyperasymptotic expansions from Loutrel and Yunes [Classical Quantum Gravity {\bf 34} 044003 (2017)]. We discuss various features of the code and study the evolution of the orbital and spin-precession parameters of eccentric, precessing binary black holes. 
In particular, we study the dependence of the spin morphologies on eccentricity, where we find that the transition point from one spin morphology to another can depend nonmonotonically on eccentricity, and the fraction of binaries in a given morphology at a given point in the evolution of a population depends on the instantaneous eccentricity.
\end{abstract}

\maketitle

\section{Introduction}
\label{sec:intro}

The LIGO and Virgo gravitational-wave (GW) detectors~\cite{advLigo2015,advVirgo2015} have so far observed about a hundred binary black holes (BBHs)~\cite{KAGRA:2021vkt, Nitz:2021zwj, Olsen:2022pin,Mehta:2023zlk,Wadekar:2023gea,LIGOScientific:2024elc}, providing initial constraints on the population of BBHs, and thus on possible formation channels for these objects~\cite{KAGRA:2021duu}. Some BBHs have been claimed to have eccentricity~\cite{Romero-Shaw:2021ual,Romero-Shaw:2022xko,Romero-Shaw:2020thy,Gayathri:2020coq,Gupte:2024jfe,Planas:2025jny} and some to have spin precession~\cite{KAGRA:2021vkt,Hannam:2021pit,Payne:2022spz,Macas:2023wiw} but the only one of these that includes both eccentricity and spin precession is \cite{Gayathri:2020coq},\footnote{A recent analysis~\cite{Morras:2025xfu} of the neutron star-black hole binary GW200105\_162426~\cite{LIGOScientific:2021qlt,KAGRA:2021vkt} reported evidence for orbital eccentricity using an inspiral-only BBH waveform model~\cite{Morras:2025nlp} that incorporates both eccentricity and spin precession.} which analyzes the high-mass signal GW190521~\cite{LIGOScientific:2020iuh,LIGOScientific:2020ufj,LIGOScientific:2021usb} using numerical relativity simulations.  This is because it was only feasible to generate the large number ($\sim 1500$) of numerical relativity simulations used to analyze GW190521 because the simulations did not need to be very long in order to analyze such a high-mass signal. To analyze signals that are not so high mass, one requires a waveform model calibrated using numerical relativity simulations, and here the first inspiral-merger-ringdown waveform models that incorporate both eccentricity and spin-precession have just recently been developed~\cite{Liu:2023ldr,Gamba:2024cvy}. The waveform models used in the analyses that claim the evidence of eccentricity assume non-precessing spins~\cite{Liu:2019jpg,Ramos-Buades:2021adz}, while the analyses that claim spin precession use waveform models with effects of spin-precession and higher-order modes~\cite{Varma:2019csw} 
but without eccentricity. Moreover, none of the analyses using waveform models can distinguish between the effects of eccentricity and spin-precession due to the possible degeneracy between the two effects~\cite{CalderonBustillo:2020xms,Romero-Shaw:2022fbf}.

The presence of either eccentricity, spin precession, or both in a GW signal hints towards the formation mechanism and environment of the binary system. For example, binaries formed through isolated formation channels in the galactic field may exhibit some eccentricity due to supernova kicks at the time of the birth of the second-formed black hole, though the emission of gravitational radiation causes the binary's eccentricity to gradually reduce during the extended inspiral~\citep{PhysRev.136.B1224,2021arXiv210812210T} until it reaches negligible eccentricity near merger (see, e.g., Fig.~5 of~\cite{Kowalska:2010qg}). However, if binaries are formed dynamically in dense stellar environments like globular clusters and nuclear star clusters, such binaries may not have sufficient time to completely shed their eccentricity before coalescence. Consequently, they can retain some eccentricity (up to $\sim1$ at $10$ Hz) when observed by ground-based detectors~\cite{Wen:2002km, OLeary:2008myb, Bae:2013fna, Antonini:2015zsa,Silsbee:2016djf,Samsing:2017xmd,Rodriguez:2017pec,Zevin:2021rtf,DallAmico:2023neb}.

Similarly, the spin magnitudes depend on how each black hole in the binary is formed. In stellar collapse, core-envelope coupling strength determines the spin magnitude~\cite{Fuller:2019sxi,Belczynski:2017gds} while black holes formed from previous mergers have a dimensionless spin magnitude around $0.7$ due to angular momentum conservation~\cite{Gerosa:2017kvu,Fishbach:2017dwv}. 
Dynamically formed binaries which experienced exchanges of components are expected to have an isotropic distribution of spin orientations with respect to the orbital angular momentum (hence exhibiting spin precession) whereas the components of isolated binaries are usually expected to inherit the collective angular momentum of their environment, leading them to have spins predominantly aligned with the orbital angular momentum (hence exhibiting very small to negligible spin precession). However, there are scenarios, discussed in~\cite{Baibhav:2024rkn}, in which isolated binaries can have very significant misalignments of the spins with the orbital angular momentum and thus significant spin precession.

Ground-based GW detectors are only able to observe the last tens of orbits of the binary evolution, so it is necessary to evolve the binary backwards in order to obtain the spin orientation and eccentricity at the time of formation and thus help elucidate the astrophysical scenario by which the binary was formed---see \cite{Johnson-McDaniel:2021rvv,Mould:2021xst,Gerosa:2023xsx,DeRenzis:2023lwa,Kulkarni:2023nes} for applications to quasicircular spin-precessing binaries and~\cite{Fumagalli:2023hde,Fumagalli:2024gko} for evolution of eccentric spin-precessing binaries both forward and backward in time with precession-averaged equations. Similarly, evolving the binary parameters forward in time and applying numerical relativity fits allows one to infer the properties of the merger remnants, useful in studying hierarchical mergers (see~\cite{Reali:2020vkf} for applications of this approach to quasi-circular precessing BBHs). For instance, while studies like~\cite{Gerosa:2017kvu} use the fact that exactly isotropic spin distributions remain isotropic, \cite{DallAmico:2023neb} finds that the distribution of spins of binaries formed in dense environments may not be exactly isotropic, since binaries that do not experience exchanges have a distribution that favors spin alignment. Thus, evolution would be necessary to study such cases.
Hence, the ability to evolve binary dynamics back and forth in time is crucial to understand the formation mechanisms and astrophysical conditions of the binary and its environment.

In this paper, we present a code to evolve eccentric, precessing BBHs from any orbital frequency to frequencies close to the merger (or vice versa) using post-Newtonian (PN) evolution equations. The spin precession is modeled using $2$PN accurate orbit-averaged precession equations from~\cite{Racine:2008qv}, thus specializing the spin-induced quadrupoles to the value for black holes. The evolution of orbital parameters such as orbital frequency, eccentricity, and periastron precession is 3PN accurate with point particle terms from~\cite{Arun:2007rg,Arun:2009mc} and leading spin-orbit and spin-spin terms from~\cite{Klein:2010ti}, including the quadrupole-monopole terms from~\cite{Klein:2018ybm}. Eccentricity enhancement functions in the fluxes use the high-accuracy super/hyperasymptotic expansions from~\cite{Loutrel:2016cdw}. The code is implemented as part of the LALSimulation package of the  LIGO-Virgo-KAGRA (LVK) collaboration's open-source library LALSuite~\cite{LALSuite} and is currently under review before it can be merged to LALSuite's master branch. A fork of LALSuite containing the unreviewed version of the code is available at~\cite{amiteshgit}. The code presented in this paper extends and corrects  the code used in Phukon~\emph{et al.}~\cite{Phukon:2019gfh}. Thus, we will highlight the differences between the new and old codes wherever applicable. There are other codes to evolve eccentric, spin-precessing, inspiraling BBHs using PN equations~\cite{Schnittman:2004vq,Cornish:2010cd,Csizmadia:2012wy,Klein:2018ybm,Ireland:2019tao,Klein:2021jtd,Arredondo:2024nsl,Morras:2025nlp}, but of these only the code associated with~\cite{Csizmadia:2012wy} (which directly evolves the PN equations without orbit averaging) is public.

In addition to evolving BBHs to study their astrophysics, the new code will also be useful in developing time-domain semi-analytic waveforms for eccentric, precessing binaries similar to the quasi-circular SpinTaylorT4 waveform, whose dynamics are described in~\cite{SpinTaylor_TechNote}, while the GW polarizations are constructed as in~\cite{Buonanno:2002fy}. Furthermore, the code will aid in twisting up aligned-spin inspiral-merger-ringdown (IMR) eccentric waveform models (such as~\cite{Nagar:2021xnh,Gamboa:2024hli,Planas:2025feq}) to construct a semi-analytic frequency domain IMR eccentric, precessing waveform, as has been done for the model from~\cite{Nagar:2021xnh} in~\cite{Gamba:2024cvy} using quasicircular precessional dynamics. This twist-up approach involves dynamical coordinate transformation of GW radiation from a noninertial frame aligned with the orbital angular momentum to the inertial frame in which the waveform is observed~\cite{Boyle:2011gg,OShaughnessy:2011pmr,Schmidt:2010it,Schmidt:2012rh}. 

In this paper, we present evolutions of example eccentric, precessing BBHs using this code. 
In particular, we replicate the three types of eccentricity evolution demonstrated in Phukon~\emph{et al.}: (i) a monotonic increase until the final separation, (ii) a monotonic decrease throughout the inspiral, and (iii) an initial decay to a minimum followed by an increase. We also study the effect of eccentricity on BBH spin morphology~\cite{Kesden:2014sla, Gerosa:2015tea} and find that the orbital separation at which binary transitions from the circulating morphology to a librating morphology 
can depend nonmonotonically on binary's initial eccentricity.  

The rest of the paper is organized as follows: In Sec.~\ref{sec:orb_elem}, we introduce the orbital parameters of an eccentric binary in the quasi-Keplerian parametrization as well as the parameters corresponding to spin precession. In Sec.~\ref{sec:ev_eq}, we give the details of the PN evolution equations for these parameters along with the initial and stopping conditions for the evolution. Sec.~\ref{sec:evols} provides results from the code for a few representative BBHs as well as BBH populations. We summarize and conclude the paper in Sec.~\ref{sec:concl}. Appendix~\ref{app:coeffs} gives the coefficients of the evolution equations, while Appendix~\ref{app:dotOmega} corrects the evolution equation for the periastron direction used in Phukon~\emph{et al.}\ and illustrates why we instead evolve the vector directly. Appendix~\ref{app:EandJ} gives the expressions for the energy and angular momentum we use in stopping conditions, and Appendix~\ref{app:example_usage} gives some examples of the use of the code. 

\textit{Conventions and Notations}: We use geometrized units ($G = c = 1$) throughout and also set $M=m^{}_1+m^{}_2 = 1$, where $M$ is the binary's total mass and $m_A^{}$ ($A \in \{1,2\}$) are its component masses, in the evolution equations and their derivation. The mass ratio and symmetric mass ratio of the binary are given by $q=m^{}_2/m^{}_1\leq1$ and $\eta=m^{}_1 m^{}_2$, respectively. We use a PN expansion parameter defined in terms of average angular frequency $\omega$ and eccentricity parameter $e_t^{}$ as $\bar{x}:= \omega^{2/3}/(1 - e_t^2)$.  We use $\omega = (1 + k)n$ for angular frequency expression, where $k$ is the periastron advance per orbit and $n$ is the mean motion of the binary~\cite{Arun:2009mc}. We denote the dimensionless spin vectors of the binary components by $\boldsymbol{\chi}^{}_{A} = \mathbf{S}^{}_{A}/m^2_{A}$, where $\mathbf{S}^{}_A$ are the spin angular momenta of the binary's components.
\section{Orbital and spin parameters}
\label{sec:orb_elem}

In this section, we introduce the parameters that are used to describe the orbital motion and spin-precession of BBHs on eccentric orbits. We first discuss the parameters describing the orbital motion, often referred to as \emph{orbital elements}.

Let us consider a binary inspiralling on an eccentric orbit with constituent masses $m^{}_1$ and $m^{}_2$,  mass ratio $q=m^{}_2/m^{}_1$, and reduced mass $\mu=m^{}_1m^{}_2$. In the Newtonian limit, one uses the \emph{Keplerian parametrization}~\cite{brouwer2013methods} to model the orbital motion of the two masses in the center-of-mass reference frame using the following parametric relations
\begin{subequations}
\label{Eq:Keplerian}
\begin{align}
r  &= a(1-e\cos u) ,\\
\phi - \phi^{}_0 &= v = 2 \arctan \left[ \left( \frac{1+e}{1-e}\right)^{1/2} \tan \frac{u}{2}\right],
\end{align}
\end{subequations}
where $r$ and $\phi$ are two parameters that describe the relative separation vector ${\mathbf r} = r(\cos \phi,\sin \phi, 0)$ of the binary components, and  $\phi^{}_0$ is the phase at periastron passage. Further, $a=(r^{}_{-} + r^{}_{+})/2$ is the semi-major axis and $e= (r^{}_{+}-r^{}_{-})/(r^{}_{+}+r^{}_{-})$ is the Keplerian eccentricity of the orbit in terms of apocenter distance $r^{}_+$ and pericenter distance $r^{}_{-}$. The angles $u$ and $v$ are called the eccentric anomaly and the true anomaly, respectively (see, e.g., Fig.~1 in~\cite{Memmesheimer:2004cv}). The time evolution of anomaly parameters appearing in Eq.~\eqref{Eq:Keplerian} can be realized by computing the eccentric anomaly $u$ from the mean anomaly $l$ at an instant $t$ by solving the following equation
\begin{equation}
\label{Eq:meanmotion}
l = n(t-t^{}_0) = u - e \sin u,
\end{equation}
where $t^{}_0$ corresponds to the time at periastron passage, and  $n=2\pi/P$ is the mean motion over a radial orbital period (periastron to periastron) $P$, which is related to the semimajor axis $a$ through Kepler's third law. Equation~\eqref{Eq:meanmotion} is referred to as {\it Kepler's equation}.

It is possible to obtain a Keplerian-type parametric solution for the orbital motion of a general-relativistic binary, thanks to the seminal work by Damour and Deruelle~\cite{AIHPA_1985__43_1_107_0,AIHPA_1986__44_3_263_0} who computed the 1PN accurate equations of motion that later led to the understanding of elliptical motion at higher PN order~\cite{Memmesheimer:2004cv,Konigsdorffer:2006zt,Arun:2007rg,Arun:2007sg,Arun:2009mc,Klein:2010ti}. Due to similarities with Kepler's equation, the PN equations describing the orbital motion of a general-relativistic binary are referred to as the \emph{quasi-Keplerian parametrization}. At a given PN order, the conservative orbital dynamics of compact binaries on eccentric orbits is specified by providing the following parametrization for the dynamical variables $r$ and $\phi$:
\begin{subequations}
\label{eq:PhasingFinalParam3PNharmonic}
	\begin{align}
		r & = a \left( 1 - e^{}_r \cos u \right) +f^{}_r(v)
		\,,
		\\
		\phi - \phi^{}_{0} & 
		= (1 + k ) v  + f^{}_\phi(v)
		\,,
		\\
		\label{phasing_eq:9c}
		\text{where} \quad
		v & = 2 \arctan 
		\left[ 
		\left( \frac{ 1 + e^{}_{\phi} }{ 1 - e^{}_{\phi} } \right)^{1/2} 
		\tan \frac{u}{2} 
		\right].
	\end{align}
\end{subequations}
In the above equations, different eccentricity parameters $e^{}_r$ and $e{}_{\phi}$ for the radial and angular variables were introduced so that the PN-accurate parametrization looks ``Keplerian'' even at higher PN orders. The quantity $k$ is related to the fraction of periastron advance per orbital revolution. In the above equations, $a$, $e^{}_r$, and $e^{}_{\phi}$ denote  the  semimajor axis, radial eccentricity, and angular eccentricity of the PN orbit, and $f^{}_r$ and $f^{}_\phi$ are some functions of the orbital energy and the angular momentum that first enter at $1$PN and $2$PN, respectively.

The following equation with a Keplerian like form links the eccentric anomaly $u$  to the mean anomaly $l = n \left( t - t^{}_0 \right)$ as
\begin{align}
	\label{eq: 3PN_KE}
	l =&\; u - e^{}_t \sin u + f^{}_t \left( u, v\right).
\end{align}
This equation introduces another eccentricity parameter, $e^{}_t$, which is usually called the time eccentricity. In this paper (and the code), we use $e^{}_t$ to define the eccentricity of the orbit which is an appropriate choice as long as binary's orbital motion is well described by PN approximation. However, in general relativity, one does not have one unique natural definition of eccentricity, though there are several proposals in the literature~\cite{Loutrel:2018ydu,Shaikh:2023ypz,Boschini:2024scu,Islam:2025oiv} to define eccentricity for general-relativistic binaries.

The parametrized solution of the orbital phase can be further decomposed into a linear part $\lambda$, often termed the mean orbital phase, and an oscillatory part $W^{}_\phi$~\cite{Damour:2004bz} as
\begin{subequations}\label{E:qk2}
\begin{eqnarray}
\phi &=& \lambda + W^{}_\phi\label{E:phi-linear} \,,\\
\dot{\lambda} &=& (1+k)n \,,\\
W^{}_\phi &=& (1+k)(v-l) + f^{}_\phi(v)\,.
\end{eqnarray}
\end{subequations}

The quasi-Keplerian parametrization for the orbital motion of spinning binaries is known up to $3$PN order for the non-spinning part~\cite{Memmesheimer:2004cv},\footnote{The 4PN order corrections to the quasi-Keplerian parametrization for orbits are derived in~\cite{Cho:2021oai}, with an approximate  treatment of certain oscillatory terms due to tail effects.} up to $3.5$PN order for the contributions from aligned spins in Arnowitt-Deser-Misner (ADM) coordinates~\cite{Tessmer:2012xr} to $3$PN in harmonic coordinates in~\cite{Henry:2023tka} and up to $2$PN order for the spin part for generically oriented spins~\cite{Klein:2010ti}. The $3$PN-accurate non-spinning expressions for $a$, $e^{}_r$, $e^{}_\phi$, $e^{}_t$, $n$, $f^{}_r$, $f^{}_\phi$, and $f^{}_t$ are given by Eq.~(20) in~\cite{Memmesheimer:2004cv} [expressed in both ADM and modified harmonic (MH) gauges] whereas the spin corrections to these quantities for generic spin orientations can be found in~\cite{Klein:2010ti}. In both gauges, the quasi-Keplerian parametrization has the same functional form, however the expressions in terms of energy and angular momentum  of the orbital elements except for $n,~k$ (i.e., $a,\, e^{}_r,\, e^{}_\phi,\, e^{}_t$) as well as for the orbital functions $f^{}_r$, $f^{}_\phi$, and $f^{}_t$ depend on choice of gauge.  In the rest of the paper, we will use the gauge invariant  Newtonian expression for the semi-major axis $a = \omega^{-2/3}$ in terms of gauge invariant average orbital angular frequency $\omega=(1+k) n$, unless stated otherwise. We use the ADM gauge expressions in our code. (While the expressions for the spin contributions in~\cite{Klein:2010ti} are in harmonic coordinates, since they start from the expressions in~\cite{Kidder:1995zr}, we do not need to perform any transformation, since the difference between ADM and MH gauge only starts at $2$PN---see, e.g.,~\cite{Arun:2007sg}. Thus, it does not affect the leading spin terms we consider.)

\begin{figure}
	\centering
	\includegraphics[width=0.495\textwidth]{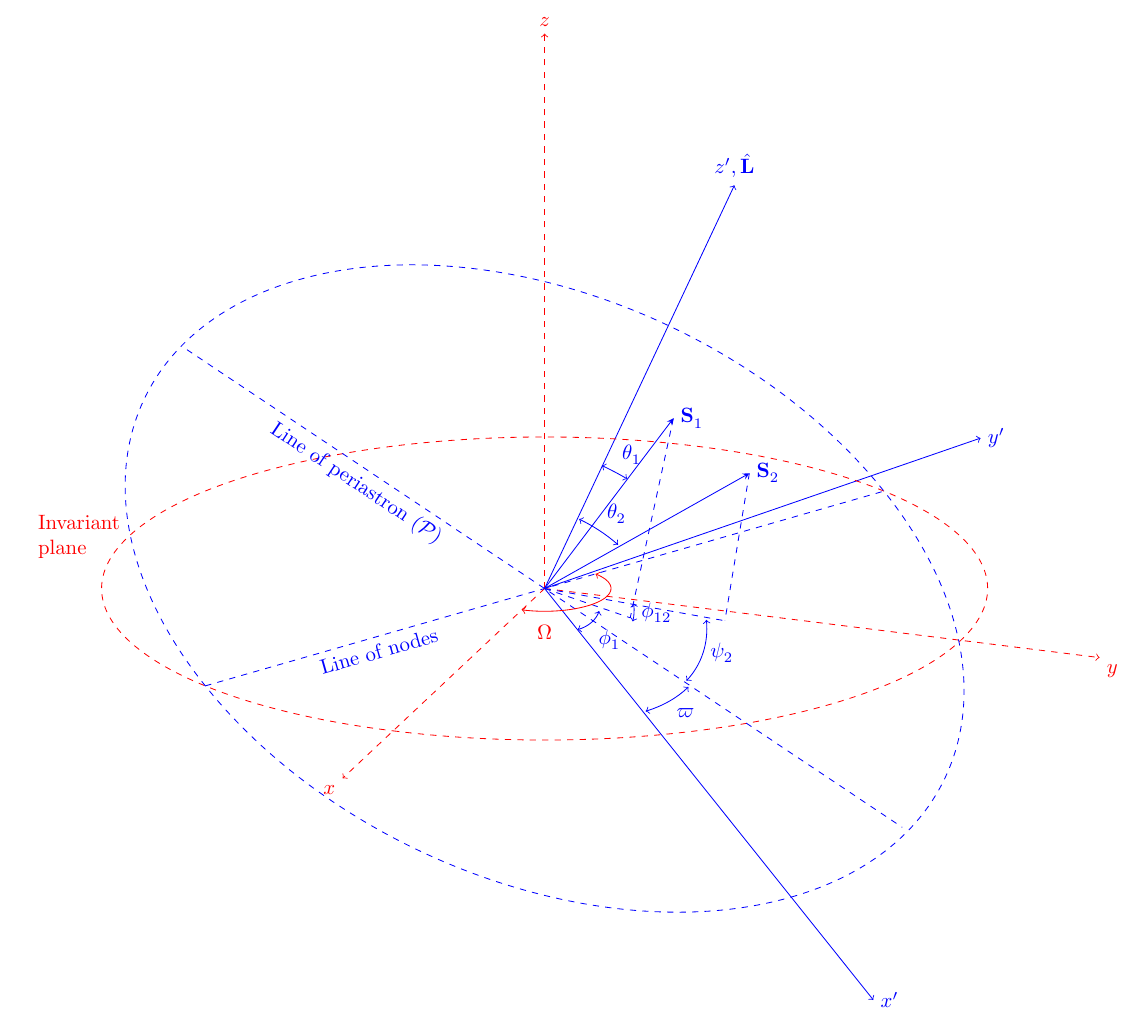}
	\caption{\label{fig:frames_angles} This figure depicts various vector quantities and the associated angles used in this study. The inertial frame is given by  $(x, y, z)$ and the invariant plane lies on the $xy$ plane of the inertial frame. The coprecessing frame $(x', y', z')$ is defined such that its $x'y'$ plane represents the instantaneous orbital plane of the binary with  $z'$-axis aligned with the  orbital angular momentum vector $\mathbf{L}$. The angles $\theta_{1,2}^{}$ are the tilt angles of $\mathbf{S}^{}_{1,2}$, while $\phi^{}_{1,2}$ is the angle between the component of $\mathbf{S}^{}_{1,2}$ in the orbital plane and the $x'$-axis.  The difference between the in-plane angles is denoted by  $\phi^{}_{12}:= \phi^{}_2 - \phi^{}_1 = \psi^{}_{2} - \psi^{}_{1}$ and $\psi^{}_{1,2}$ is the angle between $\boldsymbol{\cP}$ and the spin vector $\mathbf{S}^{}_{1,2}$. We do not show $\phi_2$ and $\psi_1$ in the figure to avoid crowding. The longitude of the line of nodes in the invariant plane is denoted by $\Omega$ and  the angle $\varpi$ measures the orientation of the line of periastron relative to the $x'$-axis, which coincides with the $x$-axis initially and then evolves due to the spin-induced precession of the orbital plane.}
\end{figure}

Let us now discuss the parameters describing the binary's spin-precession dynamics.  The relevant quantities are the spin angular momenta $\mathbf{S}^{}_1$ and $\mathbf{S}^{}_2$ of the black holes along with the binary's orbital angular momentum $\mathbf{L}$. Due to the spin-orbit and spin-spin couplings, $\mathbf{L}$, $\mathbf{S}^{}_1$, and $\mathbf{S}^{}_2$ all precess around the total angular momentum $\mathbf{J} = \mathbf{L}+ \mathbf{S}^{}_1+\mathbf{S}^{}_2$. The spin magnitudes are often defined in terms of dimensionless spin vectors $\boldsymbol{\chi}^{}_{1,2} = \mathbf{S}^{}_{1,2}/m^2_{1,2}$ with magnitudes $\chi^{}_{1,2} \in [0, 1]$. Spin orientations are described in terms of the tilt angles $\theta^{}_{1,2}\in [0,\pi]$, representing the angles between each spin angular momentum vector and the orbital angular momentum vector, along with the azimuthal angle $\phi^{}_{12} := \phi^{}_2-\phi^{}_1 \in [0, 2\pi]$ between the projections of the spin vectors onto the orbital plane. 
For eccentric, precessing binaries, two additional angles enter the dynamics, namely $\psi^{}_1$ and $\psi^{}_2$. Here $\psi^{}_1$ and $\psi^{}_2$ are defined as the angle between the line of periastron and projection of $\mathbf{S}^{}_1$ and $\mathbf{S}^{}_2$ onto the orbital plane, respectively. The line of periastron is the line joining pericentre and apocentre and precesses over time as the binary evolves, and the amount of this precession is called the longitude of periastron given as $\varpi=\lambda-l$. Finally, the line of nodes is defined as the intersection of the orbital plane with the (inertial) plane of reference, i.e., the line connecting the ascending and descending nodes. The position of the line of nodes is measured by an angle $\Omega$ with respect to the $x$-axis of the plane of reference. A pictorial description of these angles and the frames is given in Fig.~\ref{fig:frames_angles}. As discussed below, we do not use the line of nodes in our calculation, preferring instead to evolve the vector giving the periastron direction directly. However, the line of nodes was used in Phukon~\emph{et al.}, so we will discuss it while comparing with their work.

Spin precession causes the angles $\theta^{}_{1,2}$ and $\phi^{}_{12}$ to evolve over the precessional timescale. It generally takes at least several orbital cycles to complete one precessional cycle (with a larger number of orbital cycles in a precessional cycle when the binary is earlier in its evolution). For this reason, the spin dynamics can be orbit averaged~\cite{Apostolatos:1994mx}. The emission of GWs and angular momentum makes the orbit shrink (and eccentricity decay) over the radiation-reaction timescale. Additionally, for eccentric orbits, the line of periastron precesses, again generally taking at least several orbital cycles to complete a precessional cycle. (Periastron precession enters the binary dynamics at $1$PN, while spin precession starts at $1.5$PN.) Hence, the evolution of precessing compact binaries inspiraling on eccentric orbits involves four timescales:  the slowest radiation-reaction timescale, the intermediate spin-precession and the periastron-precession timescales, and the fastest orbital timescale~\cite{Apostolatos:1994mx,Damour:2004bz}. This timescale hierarchy has been extensively used in studying the orbital and spin-precession dynamics as well as in developing waveform models for such binaries including the effects of both spin precession and eccentricity~\cite{Klein:2010ti,Klein:2018ybm,Klein:2021jtd,Arredondo:2024nsl,Morras:2025nlp}.

\section{Evolution equations}
\label{sec:ev_eq}
In this section, we give the details of the differential equations used to evolve the parameters, described in the previous section, that characterize eccentric, precessing BBHs. Specifically, we evolve the binary's average orbital frequency $\omega$ and eccentricity squared $e_t^2$, as well as $\mathbf{L}$, $\mathbf{S}^{}_1$, and $\mathbf{S}^{}_2$. Additionally, we evolve the unit vector giving the position of the periastron (which we denote as $\boldsymbol{\hat{\cP}}$), since this enters into the evolution equations of some of the other quantities. We also evolve the mean anomaly $l$ and mean orbital phase $\lambda$ considered in Phukon~\emph{et al.}, for completeness, though since we are now evolving the periastron direction directly, these quantities are not used in the evolution of any other quantities. We also choose to evolve $k$ even though it is possible to compute it from the other quantities, as we do to compute its initial value (see Sec.~\ref{ssec:initial_cond}). However, we found that computing $k$ from the other quantities at each timestep gives unphysical, nonmonotonic evolution in some cases, even for nonspinning binaries, while evolving $k$ gives the expected monotonic behavior.\footnote{An example of a nonspinning binary where the analytic expression for $k$ in terms of the other quantities gives nonmonotonic evolution is one with masses of $(70, 29)M_\odot$ and an eccentricity of $0.7$ at $20$~Hz, evolving backwards to $\lesssim 5$~Hz.}

We use $\omega = (1 + k)n$, the binary's average orbital frequency (as in~\cite{Arun:2009mc}), instead of the mean motion $n$ considered in Phukon~\emph{et al.}\ so that we recover the quasicircular SpinTaylorT4 evolution~\cite{SpinTaylor_TechNote} for $e^{}_t = 0$, when we exclude the $2$PN spin contribution to the eccentricity evolution so that $e^{}_t = 0 \Rightarrow \dot{e}^{}_t = 0$. One can make this check by setting the \texttt{EccEvol2PNSpinFlag} flag to $0$ in the code (the default is $1$, where these terms are included even if $e^{}_t = 0$).
\subsection{Orbital Evolution}
\label{ssec:orb_evol_eqs}

The equations for the orbital evolution are accurate to $3$PN in the nonspinning terms (from~\cite{Arun:2009mc}) and $2$PN in the spinning terms (from~\cite{Klein:2010ti,Klein:2018ybm}), as in Phukon~\emph{et al.} However, we include the spin-induced quadrupole terms (specialized to black holes) from~\cite{Klein:2018ybm} that were not included in these equations in Phukon~\emph{et al.} The equations take the form 
\begin{widetext}
\begin{subequations}
\label{eq:evol}
\begin{align}
\dot{\omega} &= \eta \bx^{11/2}(1 - e_t^2)^2\big[\cO^{}_\text{N} + \cO^{}_\text{1PN}\bx + \cO^\text{hered, SO}_\text{1.5PN}\bx^{3/2} + \cO_\text{2PN}^\text{incl.\ SS}\bx^2 + \cO^\text{hered}_\text{2.5PN}\bx^{5/2} + \cO^\text{incl.\ hered}_\text{3PN}\bx^3\big],\\
\dot{e_t^2} &= -2\eta\bx^4(1 - e_t^2)^{3/2}\big[\cE^{}_\text{N} + \cE^{}_\text{1PN}\bx + \cE^\text{hered, SO}_\text{1.5PN}\bx^{3/2} + \cE_\text{2PN}^\text{incl.\ SS}\bx^2 + \cE^\text{hered}_\text{2.5PN}\bx^{5/2} + \cE^\text{incl.\ hered}_\text{3PN}\bx^3\big],\label{eq:sub_et2}\\
\dot{k} &= \eta\bx^4(1 - e_t^2)^{3/2}\big[\cK^{}_\text{1PN}\bx + \cK^\text{SO}_\text{1.5PN}\bx^{3/2} + \cK_\text{2PN}^\text{incl.\ SS}\bx^2 + \cK^\text{hered}_\text{2.5PN}\bx^{5/2} + \cK^{}_\text{3PN}\bx^3\big],\\
\dot{l} &= \frac{\omega}{1 + k},\\
\dot{\lambda} &= \omega,
\end{align}
\end{subequations}
\end{widetext}
where $\eta = m^{}_1m^{}_2$ is the symmetric mass ratio and $\bar{x} := \omega^{2/3}/(1 - e_t^2)$. We have denoted where spin-orbit (SO), spin-spin (SS), and hereditary (hered) terms contribute with superscripts, with ``incl.'' (for ``including'') denoting terms that also include nonspinning instantaneous contributions. The instantaneous contributions in above equations account for the linear emission of GWs from source at current retarded time, whereas hereditary terms capture tail and memory effects, that  depend over entire past of the source's dynamics~\cite{Blanchet:2013haa}.  We give the expressions for the coefficients of these equations in Appendix~\ref{app:coeffs}.

The hereditary terms include eccentricity-dependent enhancement functions, most of which cannot be computed in closed form. For the enhancement functions that cannot be computed in closed form, we give the option to either use the $O(e_t^4)$ expressions from Arun~\emph{et al.}~\cite{Arun:2009mc} used by Phukon~\emph{et al.}\ that are valid for small eccentricities, or to use the superasymptotic or hyperasymptotic expressions provided by Loutrel and Yunes~\cite{Loutrel:2016cdw} that give good accuracy even for eccentricities very close to $1$. The superasymptotic functions are constructed by optimally truncating a power series expansion in $1/(1-e_t^2)$ (either in integer or half-integer powers). The hyperasymptotic enhancement functions further improve accuracy in the small-eccentricity limit by adding corrective terms to the superasymptotic series. These corrections are given as power series containing even powers of eccentricity, $e_t^{2n}$. We give the explicit $O(e_t^4)$ expressions from Arun~\emph{et al.}\ and the equations from Loutrel and Yunes we implemented in Appendix~\ref{app:coeffs}.

The enhancement function expressions from Arun~\emph{et al.}~\cite{Arun:2009mc} that we use (and the calculations in that paper that form the basis for the Loutrel and Yunes~\cite{Loutrel:2016cdw}  expressions) set the (orbit-averaged) memory contribution to the angular momentum flux to zero. This is appropriate if the binary formed at an eccentricity very close to the eccentricities considered in the evolution;  Arun~\emph{et al.}'s   motivation was the comparison with numerical relativity simulations, which start very close to the merger. However, in general, there is a memory contribution to the angular momentum flux at Newtonian order, computed in Appendix~A of Arun~\emph{et al.} This expression involves a time integral over the binary's past history (with a nonnegative integrand, so no cancellations occur), which can be written as an integral over the binary's eccentricity evolution. It is only a significant correction to the instantaneous Newtonian energy flux if the initial eccentricity is large and the current eccentricity is not too small. Here the initial eccentricity refers to the eccentricity at the time of binary formation, where it first starts evolving solely under the influence of gravitational radiation reaction without significant influence of any additional bodies.

For instance, if the initial eccentricity is $0.8$, the maximum correction due to memory contributions is only $\sim 1\%$, and occurs when  the binary has an eccentricity of $\sim 0.4$. 
However, there can be initial eccentricities as large as  $0.9999997$ (one minus eccentricity of $3 \times 10^{-7}$) in some dynamical formation scenarios that merge within a Hubble time and still give an eccentricity as small as $\sim 0.01$ at $20$~Hz (see, e.g., the data release for~\cite{Kritos:2022ggc}), and in these cases the contribution from the memory to the angular momentum flux at $20$~Hz is still several times that of the Newtonian angular momentum flux. These calculations use Eq.~(A12) in Arun~\emph{et al.} This equation is obtained using the approximate leading-order relation between the eccentricity and semimajor axis from Peters~\cite{PhysRev.136.B1224}, which has an accuracy of $\lesssim 10\%$ and does not include the contribution of the memory to the angular momentum evolution, or any contributions from the binary's evolution when it is not evolving solely under the influence of gravitational radiation reaction. However, we are only interested in an order-of-magnitude calculation here to show the possible size of the effect---we leave more detailed calculations and implementing these corrections to future work, likely also accompanied with the implementation of higher-order spinning PN terms, which are likely to be of similar or larger importance, particularly at higher eccentricities.\footnote{There are spin-orbit terms computed through $3.5$PN~\cite{Bohe:2012mr,Bohe:2013cla} and spin-spin terms through $3$PN~\cite{Bohe:2015ana}, both for arbitrary orbits, as well as hereditary spin terms through $3$PN for eccentric binaries (computed with an expansion in eccentricity and resummed) in~\cite{Henry:2023tka}.  (This computation of hereditary terms is only for nonprecessing binaries, but this is sufficient for the application to precessing binaries through $3$PN, since the spin contributions only depend on the aligned components of the spins at $1.5$PN, and the first hereditary contributions are $1.5$PN corrections to the leading terms.) However, the quasi-Keplerian description of eccentric, precessing orbits has not been extended beyond $2$PN, so this would be necessary to extend the equations we use to higher PN order in the spins. The eccentric, precessing waveform model pyEFPE~\cite{Morras:2025nlp} incorporates the $2.5$PN and $3$PN spin-aligned terms from~\cite{Henry:2023tka} in their evolution equations, neglecting contributions from spin components perpendicular to the orbital angular momentum beyond $2$PN to achieve partial $3$PN accuracy in  spin contributions in the waveform, which they argue improves the accuracy of their model.} Even more importantly, the orbit-averaged approximation becomes less accurate for the high eccentricities for which this is the largest effect, so such studies will likely also check the accuracy of this approximation, where there are various methods available to evolve eccentric, precessing binaries without using orbit averaging, e.g.,~\cite{Csizmadia:2012wy,Loutrel:2018ydu,Ireland:2019tao,Samanta:2022yfe,Fumagalli:2025rhc}.

\subsection{Precession Equations}
\label{ssec:prec_eqs}

The precession equations are the same $2$PN leading spin-orbit and spin-spin equations as in Phukon~\emph{et al.}\ and are
\begin{subequations}
\label{eq:evo_momenta}
\begin{align}
\dot{\mathbf{S}}^{}_1 &= \frac{1}{2a^3(1 - e_t^2)^{3/2}}\left[\left(4 + 3q - \frac{3(\mathbf{S}^{}_2 + q\mathbf{S}^{}_1)\cdot\mathbf{L}}{L^2}\right)\mathbf{L} \right. \nonumber \\
&\qquad \qquad \qquad \qquad \qquad + \mathbf{S}^{}_2\Big]\times\mathbf{S}^{}_1,\\
\dot{\mathbf{S}}^{}_2 &= \frac{1}{2a^3(1 - e_t^2)^{3/2}}\left[\left(4 + 3q^{-1} - \frac{3(\mathbf{S}^{}_1 + q^{-1}\mathbf{S}^{}_2)\cdot\mathbf{L}}{L^2}\right)\mathbf{L} \right. \nonumber \\
&\qquad \qquad \qquad \qquad \qquad + \mathbf{S}^{}_1\Big]\times\mathbf{S}^{}_2,\\
\dot{\mathbf{L}} &= \boldsymbol{\omega}^{}_p\times\mathbf{L}.
\label{eq:Ldot}
\end{align}
\end{subequations}
where
\begin{subequations}
\begin{align}
\boldsymbol{\omega}^{}_p &:= \delta^{}_1\mathbf{S}^{}_1 + \delta^{}_2\mathbf{S}^{}_2, \label{eq:omegap}\\
\delta^{}_1 &:= \frac{1}{2a^3(1 - e_t^2)^{3/2}}\left(4 + 3q - \frac{3(\mathbf{S}^{}_2 + q\mathbf{S}^{}_1)\cdot\mathbf{L}}{L^2}\right),\\
\delta^{}_2 &:= \frac{1}{2a^3(1 - e_t^2)^{3/2}}\left(4 + 3q^{-1} - \frac{3(\mathbf{S}^{}_1 + q^{-1}\mathbf{S}^{}_2)\cdot\mathbf{L}}{L^2}\right),
\end{align}
\end{subequations}
with $q = m_2/m_1$.
However, instead of using $a = n^{-2/3}$ as in Phukon~\emph{et al.}, we use $a = \omega^{-2/3}$ so that we reduce to the quasicircular SpinTaylorT4 evolution~\cite{SpinTaylor_TechNote} for $e^{}_t = 0$. (These two expressions for $a$ differ starting at $1$PN.)
The magnitude of the orbital angular momentum is only needed to Newtonian order:
\<\label{eq:LNmag}
L := L^{}_N  = \eta\sqrt{(1 - e_t^2)a}.
\?
We use the Newtonian expression for $L$  to keep the precessional equations at consistent to $2$PN accuracy. The $1$PN correction to $L$ first enters the precessional dynamics at $2.5$PN order, are therefore higher-order corrections are not included here.
\subsubsection{Computation of the periastron direction}

We need to find the periastron direction in order to compute the $\psi^{}_A$ ($A\in\{1,2\}$) and $\psi$ angles appearing in the expressions for $\dot{\omega}$ and $\dot{e_t^2}$ (see Appendix~\ref{app:coeffs}). Phukon~\emph{et al.}~\cite{Phukon:2019gfh} does this by first locating the line of nodes using the angle $\Omega$ (following \cite{Konigsdorffer:2005sc}), and then finds the location of the periastron using the longitude of periastron $\varpi$. However, the expression they use to compute $\Omega$ is not correct, in general, as we show in Appendix~\ref{app:dotOmega}. Additionally, $\varpi$ is the angle of the periastron (measured in the coprecessing frame) from its initial position, not from the line of nodes. The corrected expression for $\Omega$ has singularities in some cases when $(\mathbf{S}^{}_1 + \mathbf{S}^{}_2)\,\|\,\mathbf{L}$, so we instead evolve the direction of the periastron directly, which avoids both these issues. Specifically, denoting the direction of the periastron by the unit vector $\boldsymbol{\hat{\cP}}$, we have
\<
\label{eq:omega}
\dot{\boldsymbol{\hat{\cP}}} = [\boldsymbol{\omega}^{}_p - (\boldsymbol{\omega}^{}_p\cdot\mathbf{\hat{L}})\mathbf{\hat{L}} + kn\mathbf{\hat{L}}]\times\boldsymbol{\hat{\cP}}.
\?
The $\boldsymbol{\omega}^{}_p$ part of this equation comes from the equation $\dot{\boldsymbol{\hat{\mathrm{L}}}} = \boldsymbol{\omega}^{}_p\times\mathbf{\hat{L}}$ [recall that $\boldsymbol{\omega}^{}_p$ is given by Eq.~\eqref{eq:omegap}]. This equation comes from Eq.~\eqref{eq:Ldot} upon noting that the latter equation implies that $L$ is a constant. We require $(\boldsymbol{\hat{\cP}}\cdot\mathbf{\hat{L}})^{\textstyle\cdot} = (\boldsymbol{\hat{\cP}}\cdot\boldsymbol{\hat{\cP}})^{\textstyle\cdot} = 0$, so that the evolution keeps $\boldsymbol{\hat{\mathrm{L}}}$ and $\boldsymbol{\hat{\cP}}$ orthogonal and $\boldsymbol{\hat{\cP}}$ a unit vector if they are initially. Thus, we obtain $\dot{\boldsymbol{\hat{\cP}}} = (\boldsymbol{\omega}^{}_p + \alpha\mathbf{\hat{L}})\times\boldsymbol{\hat{\cP}}$ for any $\alpha$. We fix $\alpha$ by noting that we first want to remove the component of $\boldsymbol{\omega}^{}_p$ parallel to $\mathbf{L}$, since the component along $\mathbf{L}$ just gives rotation around $\mathbf{L}$, not precession of the orbital plane [cf.\ Eqs. (71, 72) in~\cite{Buonanno:2002fy}]. Once we do that, we have an equation that gives the evolution of the initial periastron location under the (spin-induced) precession of the orbital plane. We then add the $kn\mathbf{\hat{L}}$ term to include the precession of the periastron due to eccentricity (which occurs in the orbital plane), recalling that $\varpi$ gives the amount of periastron advance and $\dot{\varpi} = kn$.

\subsubsection{Computation of $\psi$s}

We can now compute the $\cos 2\psi^{}_A$ and $\cos 2\psi$ that appear in the expressions for $\dot{\omega}$ and $\dot{e_t^2}$ using the $\boldsymbol{\hat{\cP}}$ vector. However, instead of performing rotations using Euler angles to place the spins in the appropriate frame to compute the angles, as in Phukon~\emph{et al.}, which can lead to difficulties when the Euler angle parameterization becomes singular, we instead compute the cosines directly using inner products, noting that $\cos 2\alpha = 2\cos^2\alpha - 1$ for any $\alpha$. To compute the $\cos\psi$s, we first project the spins into the plane of the orbit, giving $\boldsymbol{\chi}_A^\perp = \boldsymbol{\chi}^{}_A - (\boldsymbol{\chi}^{}_A\cdot\mathbf{\hat{L}})\mathbf{\hat{L}}$, so we have
\begin{subequations}
\begin{align}
\cos\psi^{}_A &= \frac{\boldsymbol{\chi}_A^\perp\cdot\boldsymbol{\hat{\cP}}}{\|\boldsymbol{\chi}_A^\perp\|},\\
\cos\psi &= \frac{\mathbf{s}^\perp\cdot\boldsymbol{\hat{\cP}}}{\|\mathbf{s}^\perp\|},
\end{align}
\end{subequations}
where $\mathbf{s}^\perp = m^{}_1\boldsymbol{\chi}_1^\perp + m^{}_2\boldsymbol{\chi}_2^\perp$. To prevent division by zero in the code, we set $\cos\psi^{}_A$ and $\cos\psi$ to zero if the denominator in the above expressions is zero. This is permissible because $\cos\psi^{}_A$ and $\cos\psi$ only appear multiplied by quantities that include the squares of these denominators [see Eq.~\eqref{eq:sigma}].

\subsection{Initial Conditions}
\label{ssec:initial_cond}

Recall that the periastron advance quantity $k$ is not independent of the other quantities, but we evolve it in order to prevent unphysical behavior, as discussed previously. We thus need to compute the initial value of $k$ from the initial conditions for $\omega$, $e_t^2$, and the spins. The $3$PN expression with no spin contributions is obtained from Eq.~(6.2b) and Eq.~(6.5) in Arun~\emph{et al.}~\cite{Arun:2009mc}, the leading spin-orbit contribution comes from Eq.~(26) in Klein and Jetzer~\cite{Klein:2010ti}, and the leading spin-spin contribution from Eq.~(A5c) in Klein~\emph{et al.}~\cite{Klein:2018ybm}, recalling that we are specializing to black hole spin-induced quadrupoles. We thus have
\<
\begin{split}
\label{eq:ini_k}
k &= 3 \bx - \beta(4,3)\bx^{3/2}\\
&\quad + \frac{1}{4}\left[54 - 28\eta + (51-26 \eta )e_t^2 + 6\gamma^{}_1\right]\bx^2\\
&\quad + \frac{1}{4}\biggl\{210 + \left(\frac{123 \pi^2}{8}-625\right)\eta + 28 \eta^2\\
&\quad + \left[567 + \left(\frac{123 \pi ^2}{32}-816\right)\eta + 160\eta^2\right]e_t^2\\
&\quad + \left(78  - 55\eta + \frac{65}{2}\eta^2\right)e_t^4\\
&\quad + \left[60 - 24\eta + (120-48 \eta )e_t^2\right]\sqrt{1-e_t^2}\biggr\}\bx^3,
\end{split}
\?
where
\begin{subequations}
\label{eqs:betagamma}
\begin{align}
\beta(a, b) &:= \Big[a\Big(\BS^{}_1  + \BS^{}_2\Big) + b\Big(q\BS^{}_1 + \frac{\BS^{}_2}{q}\Big)\Big]\cdot\BhL,\\
\gamma^{}_1 &:= \frac{3(\Bs\cdot\BhL)^2 - \|\Bs\|^2}{2}.
\end{align}
\end{subequations}
Here $\Bs := \BS^{}_1/m^{}_1 + \BS^{}_2/m^{}_2$.

Since this initial condition does not agree exactly with the result obtained from evolving the equations to this frequency, we also give the option in the code of passing the initial value of $k$, as discussed in Appendix~\ref{app:example_usage}, but the default is to use the computation given above for the initial value of $k$. In particular, if one wants to reproduce the initial values (up to numerical accuracy) when evolving forward and then backward to the initial frequency, or vice versa, one has to pass the final value of $k$ from the first evolution as the initial value for the second evolution.

We also check that the initial conditions do not lead to unphysical results, i.e., that the binary is bound (its energy is negative) and the PN expression for the magnitude of its orbital angular momentum and $k$ are positive, and the time derivatives of both the energy and angular momentum are negative.

\subsection{Evolution and Stopping Conditions}
\label{ssec:stopping_cond}

The evolution of differential equations presented above is carried out using the GNU Scientific Library implementation of the fourth-order adaptive Runge-Kutta-Fehlberg $(4,5)$ solver (RKF45), with additional Hermite interpolation as implemented in LALSuite. We set both the absolute and relative tolerances to $10^{-12}$. The evolution requires initial values of $\boldsymbol{\chi}^{}_1$, $\boldsymbol{\chi}^{}_2$, $\mathbf{\hat{L}}$, and $e^{}_t$ (along with $m^{}_1$, $m^{}_2$) and returns the evolved values of these quantities, as well as $\omega$, $l$, $\lambda$, and $\boldsymbol{\hat{\cP}}$. One can evolve either forward or backwards and there is an option for either outputting the full time series or just the final values. The specific usage of the code is given in Appendix~\ref{app:example_usage}.

\begin{table}
\caption{\label{tab:stopcond} The binary's evolution stops if any of these conditions is true, and our code prints the stopping code if debug output is turned on. All but stopping code 1029 correspond to an unexpected termination of the evolution.}
\begin{tabular}{cc}
\hline\hline
Condition & Stopping Code\\
\hline
$\dot{\varepsilon}<0$    & 1025 \\
$\ddot{\omega}\le0$  & 1026 \\
$\omega$ is a nan     & 1028 \\
$f^{}_{\rm GW}$ is past $f^{}_{\rm end}$ & 1029 \\ 
$f^{}_{\rm GW}<0$    & 1030 \\
$\bar{x}\ge1$  & 1031 \\
$\dot{L}>0$    & 1032 \\
$e_t^2 \notin [0,1)$ & 1033 \\
$\varepsilon<0$    & 1034 \\
$L<0$    & 1035 \\
$k<0$    & 1036 \\
\hline\hline
\end{tabular}
\end{table}

We check that the PN equations are not giving obviously wrong results by performing various checks at each timestep, listed in Table~\ref{tab:stopcond}. (The code gives the option to evolve forward as far as possible until one of these stopping conditions is triggered.) The first six checks are the same as in the quasicircular SpinTaylor code in LALSuite, except with the replacement of the orbital velocity by $\bar{x}$, and have the same stopping codes. (Stopping code 1027 was already assigned in the quasicircular SpinTaylor code, but is not used for any test there, so we do not use this number in our code.) We check if $\bar{x} \geq 1$ because the orbital evolution equations \eqref{eq:evol} we consider are a series in $\bar{x}$, so if $\bar{x} \geq 1$ we are in a regime where these equations are surely not trustworthy. The other checks that are the same as for the SpinTaylor code are checking if the binary's energy is increasing ($\dot{\varepsilon}<0$, where the energy quantity $\varepsilon$ is given in Appendix~\ref{app:EandJ}) and if the rate of change of orbital frequency is not increasing with time ($\ddot{\omega}\le0$), as well as checks that the frequency has not passed the desired ending frequency $f^{}_{\rm end}$, is positive, and is not a nan. The energy is expected to decrease, due to the emission of GWs, and $\ddot{\omega}$ is expected to increase as one approaches merger (this is true for any frequency at Newtonian order, but is no longer guaranteed to be true at high frequencies starting with the 1PN corrections or the introduction of spin-orbit terms even for $e^{}_t = 0$). The $\ddot{\omega}\le0$ check is implemented by comparing the $\dot{\omega}$ values at consecutive timesteps. Since the angular momentum flux is not simply related to the energy flux in the eccentric case, we also check whether the magnitude of the binary's orbital angular momentum is increasing ($\dot{L}>0$, where the orbital angular momentum magnitude $L$ is given in Appendix~\ref{app:EandJ}). We also check if the binary has become unbound ($\varepsilon<0$); the PN expression for the magnitude of the orbital angular momentum or the periastron precession quantity has become negative ($L<0$ or $k<0$); and if $e_t^2$ is outside of the allowed range of $[0,1)$.

The ``GW frequency'' we use to determine the starting and ending frequencies of the evolution is just given by $f_{\rm GW} = \omega/\pi$, so it gives the $m = 2$ GW frequency in the quasicircular case. It presumably gives some orbit-averaged GW frequency in the eccentric case, but it would be necessary to compute the waveform to check how this frequency compares to, e.g., the average frequencies considered in~\cite{Shaikh:2023ypz}, so we leave this for future work. Also, to be explicit, the check that $f^{}_{\rm GW}$ is past $f^{}_{\rm end}$ corresponds to
\<
\begin{split}
f^{}_{\rm GW} > f^{}_{\rm end}, &\;\;\text{if } f^{}_{\rm end} >  f^{}_{\rm start},\\
f^{}_{\rm GW} < f^{}_{\rm end}, &\;\;\text{if } f^{}_{\rm end} < f^{}_{\rm start}.
\end{split}
\?

\section{Evolution of Binary Black Holes}
\label{sec:evols}
In this section, we present the time evolution of a few representative eccentric, precessing BBHs using our code, including checking the effect of different choices for the enhancement functions and the order of eccentricity corrections used in the hyperasymptotic expressions for the enhancement functions. For all the analyses in this section, the timestep for the interpolation is fixed to $\Delta T=1/16384$~s which allows us to obtain finely sampled timeseries for all our output variables, while also minimizing the disk storage required to save the full timeseries outputs. 

\subsection{Effect of Eccentricity and Spins}
\label{ssec:quasicirc_comp}

We consider a set of BBHs with total mass $M=20M_\odot$, mass ratio $q\in\{1/1.2, 1/4\}$ and spin magnitudes $\chi^{}_{1,2}\in\{0.5, 0.95\}$ [i.e., we have the four combinations $(\chi^{}_1, \chi^{}_2)=(0.5, 0.5), (0.5, 0.95), (0.95, 0.5), (0.95, 0.95)$]. The initial eccentricity $e^{}_0$ of all these BBHs is given at $f^{}_{\rm start}=10$ Hz and we choose four values: $e^{}_0\in\{0, 0.1, 0.5, 0.7\}$. These choices result in $32$ BBHs with different mass ratios, spin magnitudes, and initial eccentricities. The initial spin angles for all these $32$ BBHs are kept the same and are randomly chosen once from their respective domains. Specifically, at $f^{}_{\rm start}$ we have $\theta^{}_1=0.9938$~rad, $\theta^{}_2=0.6211$~rad, $\phi^{}_1=3.6118$~rad and $\phi^{}_2=1.3118$~rad. We evolve all BBHs forward as far as possible using the highest PN orders available in our code ($3$PN accurate nonspinning terms and $2$PN accurate spinning terms). For each binary, we select all three types of enhancement functions as discussed in Sec.~\ref{ssec:orb_evol_eqs}, i.e., the enhancement functions from Arun~\emph{et al.}~\cite{Arun:2009mc}, and the ones from Loutrel and Yunes~\cite{Loutrel:2016cdw} (see Appendix~\ref{app:enhancement}). 
Evolution of all but two binaries ended when the time derivative of the energy became positive (stopping code 1025). Those two binaries were evolved using Arun~\emph{et al.}~\cite{Arun:2009mc} enhancement functions, the first one with $q=1/1.2$, $(\chi^{}_{1},\chi^{}_{2})=(0.95,0.5)$ and the second with $q=1/4$, $(\chi^{}_{1},\chi^{}_{2})=(0.5,0.95)$, where $e^{}_{0}=0.7$ for both. Both produced the stopping code 1032 ($\dot{L}>0$).
 
\begin{figure*}
\includegraphics[width=.99\textwidth,height=14cm]{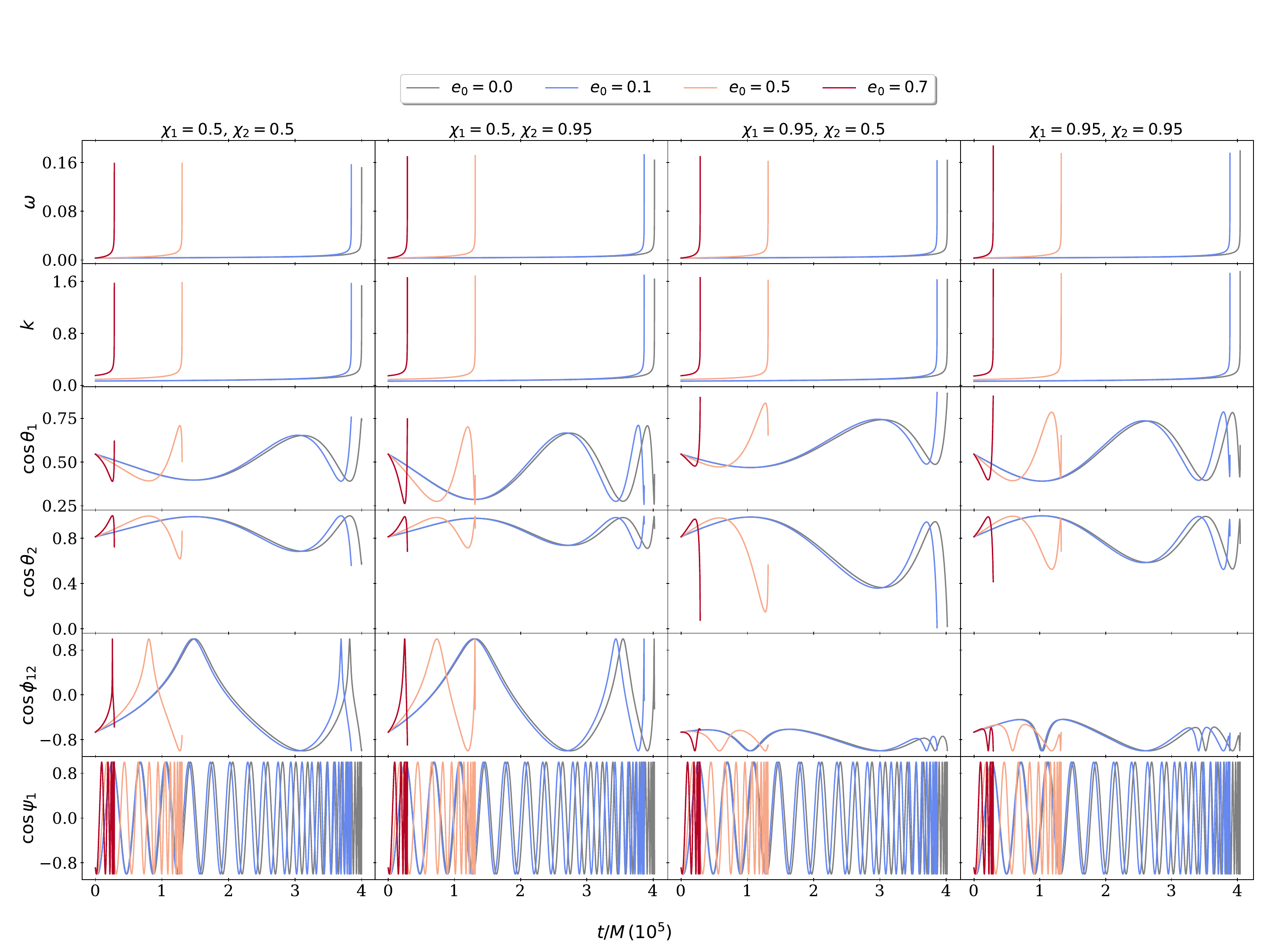}
\caption{The figure shows the time evolution of orbital ($\omega$, $k$) and spin ($\cos\theta^{}_{1}$, $\cos\theta^{}_{2}$, $\cos\phi^{}_{12}$, $\cos \psi^{}_{1}$) parameters of binaries with $q=1/1.2$ using the hyperasymptotic enhancement function from Loutrel and Yunes~\cite{Loutrel:2016cdw}. Different colors depict different values of the initial eccentricity $e^{}_0$ at $f^{}_{\rm start}=10$ Hz.}
\label{fig:binaryevolution}
\end{figure*}

Figure~\ref{fig:binaryevolution} shows the time evolution of $\omega$, $k$, $\cos\theta^{}_{1}$, $\cos\theta^{}_{2}$, $\cos\phi^{}_{12}$, and $\cos\psi^{}_{1}$ for BBHs with $q=1/1.2$ and hyperasymptotic enhancement function from Loutrel and Yunes~\cite{Loutrel:2016cdw}. We do not show the results for binaries with $q=1/4$ and other enhancement functions since the evolutions of binary parameters are very similar to what is shown in Fig.~\ref{fig:binaryevolution}. We find the expected faster evolution for higher initial eccentricities (since eccentric binaries lose energy and angular momentum more rapidly than quasicircular binaries).
The evolution of $\cos\phi^{}_{12}$ is very different in the first two and the last two columns, since those binaries are in different spin morphologies~\cite{Kesden:2014sla,Gerosa:2015tea}. The binaries in the two right-most columns are in the spin morphology where $\phi^{}_{12}$ librates around $\pi$, while those in the left-most two columns are in the more common spin morphology where $\phi^{}_{12}$ circulates over its entire range of $[0,2\pi]$. (This difference in spin morphologies is just a consequence of our randomly sampled parameters, and was not chosen intentionally.) The oscillation frequency of $\cos \psi^{}_{1}$ increases as the binary approach merger due to the increasing periastron precession frequency. The evolution of $\cos \psi^{}_{2}$ is very similar to that of $\cos \psi^{}_{1}$ (indeed, $\psi^{}_2 = \psi^{}_1 + \phi^{}_{12}$) and hence is not shown in Fig.~\ref{fig:binaryevolution}.

In the $e^{}_0 = 0$ case, we also check the difference in the evolution due to the $2$PN spin terms in the eccentricity evolution equation that cause the eccentricity to increase rather than staying zero throughout the evolution. If we exclude those terms from the eccentricity evolution equations, then we find small differences, with the largest difference (normalized to the difference between minimum and maximum values from the evolution) of $0.1781$ in $\cos\psi^{}_{2}$ for the binary with $q=1/1.2$, $(\chi^{}_{1}, \chi^{}_{2})=(0.95,0.95)$, using Loutrel and Yunes's hyperasymptotic enhancement function. The final eccentricity reached for this binary is $e^{}_{t}=0.0095$ when the $2$PN spin terms in the eccentricity evolution were turned on. Across all 8 binary configurations with $e^{}_{0}=0$, the largest final eccentricity is $e^{}_{t}=0.0193$ for the binary with $q=1/1.2, (\chi^{}_{1},\chi^{}_{2})=(0.95,0.95)$ with Loutrel and Yunes's superasymptotic enhancement function.

\begin{figure*}
	\includegraphics[width=.99\textwidth,height=6.5cm]{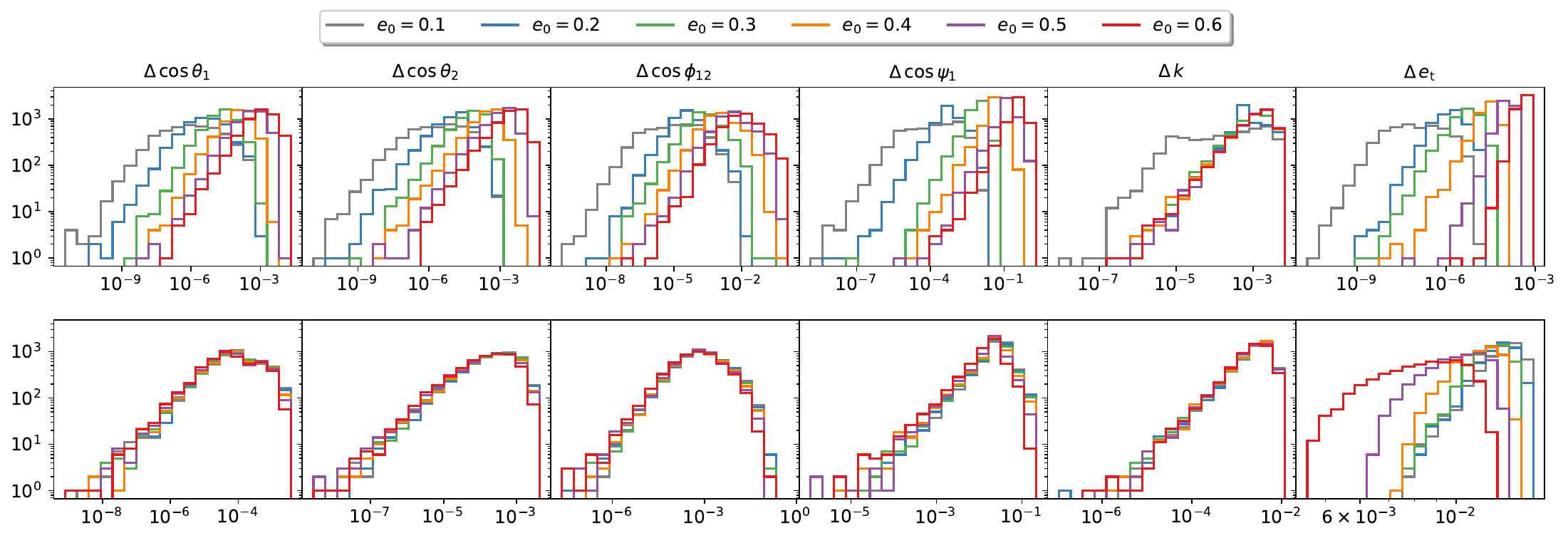}
	\caption{The figure shows the histogram of absolute differences in spin angles ($\cos\theta^{}_{1}$, $\cos\theta^{}_{2}$, $\cos\phi^{}_{12}$, and $\cos \psi^{}_{1}$) and orbital parameters ($k$ and $e^{}_{t}$) for binaries evolved with different enhancement functions. The top panel shows the differences between evolutions using the hyperasymptotic enhancement functions from Loutrel and Yunes~\cite{Loutrel:2016cdw}, and the $O(e_t^4)$ enhancement functions from Arun~\emph{et al.}~\cite{Arun:2009mc} with their respective highest eccentricity order. The bottom panel compares the hyperasymptotic enhancement functions with the superasymptotic enhancement functions. Different colors depict different values of the initial eccentricity $e^{}_0$ at $f^{}_{\rm start}=10$ Hz. It is evident that higher eccentricities correspond to larger differences in the top panel since the hyperasymptotic enhancement functions are more accurate for large eccentricities than the ones from Arun~\emph{et al.} In the bottom panel, we observe that the differences, especially in $e^{}_{t}$, are larger at smaller initial eccentricities. This is expected, as the hyperasymptotic corrections to the superasymptotic enhancement functions are more important for lower eccentricities.}
	\label{fig:enfunc_comparison_hist}
\end{figure*}

\subsection{Effect of Accuracy of Enhancement Functions}
\label{ssec:ecc_order_comp}

As mentioned in Sec.~\ref{ssec:orb_evol_eqs}, the Loutrel and Yunes enhancement functions come in two forms: superasymptotic and hyperasymptotic, where the hyperasymptotic functions improve the accuracy of the superasymptotic functions by adding a polynomial in $e_t^2$. The accuracy increases as one includes higher and higher terms in this polynomial. We thus consider the effects of using different numbers of terms: We refer to a case where terms up to $e_t^{2n}$ are included as having \emph{eccentricity order} $2n$. The expressions of hyperasymptotic functions provided in~\cite{Loutrel:2016cdw} contain terms up to eccentricity order $20$. We also compare the different enhancement functions with varying eccentricities to see when they can produce significant differences in the evolution of binary parameters.%

We choose a relatively low total mass $M=20M^{}_\odot$, to get a relatively long evolution so that we could capture the highest differences in the timeseries. We choose spin angles arbitrarily as $\theta^{}_{1}=0.5$~rad, $\theta^{}_{2}=3.0$~rad, $\phi^{}_{1}=5.1$~rad, and $\phi^{}_{2}=3.7$~rad at $f^{}_{\rm start}=10$~Hz. We sample $q$ uniformly from $[0.05, 1]$, and  $\chi^{}_{1}$ and $\chi^{}_{2}$ from $[0,1]$. We consider the initial eccentricities, $e^{}_0=\{0.1, 0.2, 0.3, 0.4, 0.5, 0.6, 0.7, 0.9, 0.99\}$, and evolve binaries forward as far as possible using the highest PN orders available in our code ($3$PN accurate nonspinning terms and $2$PN accurate spinning terms). For each $e^{}_0$ value, we evolve binaries considering all eccentricity orders of hyperasymptotic enhancement functions (ten of them) from order $2$ (i.e., up to $e_t^2$ terms and least accurate) to order $20$ (i.e., up to $e_t^{20}$ terms and most accurate). We compare the evolutions using enhancement functions truncated at orders $2$ and $20$. For the $e^{}_{0}=0.1$ case, we find very small differences between the time series of various binary parameters---the maximum absolute difference among all parameters was around $4\times 10^{-2}$ for $\cos\psi^{}_{2}$. For $e^{}_{0}=0.6$, the maximum absolute difference approached $5.4\times10^{-2}$ for $\cos\psi^{}_{2}$. For higher eccentricities ($e^{}_{0}\in \{0.7,0.9,0.99\}$), we found that all of the binaries' initial conditions produced $\bar{x} \ge 1$, and thus the evolution did not start, since the PN equations we are using are definitely not trustworthy in such situations. 

We also investigate the effect of initial eccentricity $e^{}_0$ on binary evolution while using different enhancement functions, and the results are presented in Fig.~\ref{fig:enfunc_comparison_hist}. The top panel of Fig.~\ref{fig:enfunc_comparison_hist} shows the difference in final values of $\cos \theta^{}_1$, $\cos \theta^{}_2$, $\cos \phi^{}_{12}$, $\cos \Psi^{}_1$, $k$, and $e^{}_{t}$, when binaries were evolved using either the Loutrel and Yunes hyperasymptotic enhancement functions or the Arun~\emph{et al.}\ expressions for the enhancement functions using their most accurate form, whereas the bottom panel shows the same comparing the hyperasymptotic and superasymptotic enhancement functions. We observe that there are relatively significant differences between the results with the hyperasymptotic and Arun~\emph{et al.}\ enhancement functions for larger eccentricities (for $e^{}_0=0.6$, differences of up to $1.12$ in $\cos \phi^{}_{12}$ and $1.99$ in $\cos \psi^{}_{1}$). For smaller eccentricities, such as for $e_{0}=0.1, 0.2$, the largest maximum absolute differences between the results with these two enhancement functions occur for $\cos \psi_{1}$, with values of $0.012$ and $0.020$, respectively, while the smallest maximum differences occur for $e_t$, with values of $8.7 \times 10^{-6}$ and $2.1 \times 10^{-5}$, respectively. These sorts of results are expected, since the Arun~\emph{et al.}\ enhancement functions are a power series truncated at $O(e_t^4)$, while the hyperasymptotic enhancement functions include power series in $1/(1-e_t^2)$ that accurately capture the rapid growth of the enhancement functions for large eccentricities.

On the other hand, the difference between the hyperasymptotic and superasymptotic evolutions is generally very small, as expected, since the binary evolution with higher order $O(e_t^{20})$ terms in hyperasymptotic enhancement functions have quite small differences as compared to evolution with just $O(e_t^2)$ terms, as we have seen above. However, we do find slightly larger differences between the hyperasymptotic and superasymptotic evolutions in some cases, with a maximum difference of $0.53$ in $\cos\phi^{}_{12}$ and $0.22$ in $\cos\psi^{}_{1}$. The cases which correspond to the top $30$ largest differences in $\cos \phi^{}_{12}$ have mass ratios smaller than $0.31$, and thus the tilt angles themselves at the end of the evolution are not very different from the initial values, which leads to a relatively small in-plane spin for the secondary. In fact, we find that in the case that gives the largest difference in $\cos\phi^{}_{12}$, the final value of $\sin\theta^{}_2$ is $\sim 10^{-3}$. Thus, the in-plane spin component is very small, leading to large differences in $\cos\phi^{}_{12}$ from small differences in the spin components, though this is a finely tuned case. Additionally, as expected, the differences between the hyperasymptotic and superasymptotic evolutions are largest for smaller eccentricities ($e^{}_{0}\in \{0.1, 0.2 \}$), since these are the eccentricities for which the hyperasymptotic corrections are most important. All these results demonstrate that one should use the hyperasymptotic enhancement functions to obtain accurate results for binaries with larger eccentricities, though the Arun~\emph{et al.}\ enhancement functions can be sufficient for sufficiently small eccentricities.

\section{Interplay of Eccentricity and Spin Dynamics in Radiation Reaction Timescale}
\label{sec:ecc_vs_dynamics}

We now delve into the interplay of eccentricity and spin dynamics in eccentric, precessing BBHs, focusing on the binary's evolution under radiation reaction. The evolution of $e_t$ can be complex for precessing compact binaries, including non-monotonic evolution and even overall eccentricity growth at the end of inspiral~\cite{Klein:2010ti}. The spin dynamics of precessing binaries are intriguing in their own right due to the rich dynamical phenomena that shape the evolution of a precessing binary system~\cite{Apostolatos:1994mx}. Spin orientations in precessing compact binaries evolve on two distinct timescales: the faster precession timescale and the slower radiation reaction timescale. In particular, the spin morphology, a feature of the spin dynamics of binaries with both spins nonnegligible, is a property of the conservative precessional evolution, but binaries can transition from one morphology to another as they evolve under radiation reaction~\cite{Kesden:2014sla,Gerosa:2016sys}.  
This behavior makes spin morphologies a useful alternative to spin orientations for constraining the binary's state near merger based on  its initial state at formation in precessing binaries or vice versa~\cite{Gerosa:2015tea,Reali:2020vkf}. It is thus crucial to understand the phenomenology of the evolution of spin-morphologies and eccentricity driven by GW emission, as both features bear signatures of the astrophysical processes of formation of these systems.

\subsection{Eccentricity evolution}
\label{ssec:ecc_evol}

Phukon {\it et al.}~\cite{Phukon:2019gfh} studied evolutions of eccentricities and spin morphologies on the radiative reaction timescale for a set of binaries evolving from a semimajor axis of $a=1000 M $ to $a=10 M$. They illustrated three distinct types of eccentricity evolution in generic binaries and also found that the evolution of spin morphologies in the eccentric case is qualitatively similar to the quasicircular case.  A limitation of this study was that it was conducted in limited parameter space and used less accurate equations describing the evolution of binaries.  Recently,  Fumagalli and Gerosa~\cite{Fumagalli:2023hde} have developed a PN framework to study eccentric, precessing binaries in radiation reaction timescale. Several other studies have also investigated   the interplay of eccentricity and spin dynamics in eccentric, precessing binary systems~\cite{Schnittman:2004vq,Klein:2010ti,Ireland:2019tao}. However,  these studies employed somewhat incomplete prescriptions, such as the use of  Keplerian orbits or the effects of the leading-order radiation reaction effects on binaries.  In our exploration of the evolution of eccentricity, spin dynamics, and their interplay, we use a wider parameter space than~\cite{Phukon:2019gfh}, though with the same range of orbital separations, while employing the most accurate equations currently available to describe binary evolution.

We use our code to study the eccentricity evolution of eccentric, precessing BBHs. In this study, we use the highest PN orders available in our code ($3$PN accurate nonspinning terms and $2$PN accurate spinning terms), as in the previous section. We also use the highest order hyperasymptotic enhancement functions. We evolve $50000$ random samples of  binaries, distributed uniformly in  mass ratio $q$  covering the  range $q \in [1/20,\,1)$, in black hole spin magnitudes in the range $\chi^{}_{1,2} \in (0, 0.99]$. In contrast, Phukon {\it et al.}~\cite{Phukon:2019gfh} considered a grid of  smaller number of fixed values of $q$, $\chi_1^{}$, and $\chi_2^{}$.   The polar $\theta^{}_{1,2}$ and azimuthal $\phi^{}_{1,2}$  angles of spins are chosen such that spin unit vectors are uniformly distributed over the surface of a unit sphere.  The initial eccentricities of the binaries are defined at $a=1000M$, i.e., $e_0^{}:=e^{}_t(1000M)$ and uniformly distributed with the upper bound $ e_0^{}=0.9$. Here $a = (\pi M f^{}_\text{GW})^{-2/3}M$ is the Newtonian semimajor axis. For computational efficiency, we evolve each binary from the initial orbital separation $a=1000M$ to the orbital separation corresponding to GW frequency 10 Hz using a timestep of  $\Delta T = 1/64$~s then continued the evolution till $a=10M$ with a finer timestep of $\Delta T = 1/8192$~s.

\begin{table}
\caption{\label{tab:BinarySimParams}Distributions of parameters of eccentric binary population at orbital separation $a=1000M$.  In the table, $\mathrm{U}$ denotes uniform distribution in the range ($x, y$) for respective parameters. For the eccentricity distribution, a fifth of the binaries use the residual eccentricities, as described in the text.}
\begin{tabular}{c@{\quad}c}
\hline
\hline
Parameter & Distribution  \\
\hline
Mass ratio $(q=m_2^{}/m_1^{})$ & $\mathrm{U} [0.05,\,1)$\\
\multirow{2}{*} { Eccentricity ($e^{}_0$ at $a=1000 M$) } & $\mathrm{U}[10^{-5},\,0.9]$  \\
& \& $\{e_{\rm res}^{1000}(q, \boldsymbol{\chi}_1, \boldsymbol{\chi}_2)$\}\\
Primary spin magnitude ($\chi_1^{}$) & $\mathrm{U}(0,\,0.99]$ \\
Secondary spin magnitude ($\chi_2^{}$) & $\mathrm{U}(0,\,0.99]$ \\
Cosine of primary tilt $( \cos \theta_1^{})$ & $\mathrm{U} (-1, 1)$\\
Cosine of secondary tilt $( \cos \theta_2^{})$ & $\mathrm{U} (-1, 1)$\\
Azimuthal angle of primary spin $(\phi^{}_1)$ & $\mathrm{U}[-\pi,\pi]$ \\
Azimuthal angle of secondary spin  $(\phi^{}_2)$ & $\mathrm{U} [-\pi,\pi]$ \\
\hline
\hline
\end{tabular}
\end{table}%

We use a mixture distribution for the eccentricities of binaries at $a=1000M$. Of the  initial eccentricities at $a=1000M$, $40000$ samples are drawn from a uniform distribution in the range $[10^{-5}, 0.9]$. The other set of $10000$ eccentricities are taken to be the residual eccentricities $e_{\rm res}^{1000}$, computed at the initial separation ($a=1000M$) for each of the random binaries with masses and spin parameters drawn from distributions given above. 
The residual eccentricity $e_{\rm res}^{}$ is the lower bound of orbital eccentricity for a given precessing binary that comes about due to the 2PN spin-spin contributions to the evolution of $e^{}_t$, as discussed in \cite{Klein:2010ti,Klein:2018ybm}.\footnote{This effect is present in the orbit-averaged evolution and even for single-spin cases and is separate from the effect of the conservative 2PN spin-spin dynamics where circular orbits only exist in an orbit-averaged sense for binaries with both spins nonzero (see, e.g.,~\cite{Kidder:1995zr}). In particular, as discussed in~\cite{Buonanno:2010yk}, the conservative spin-spin effect gives modulations at twice the orbital frequency, while eccentricity affects the orbit at the orbital frequency.} One can obtain the minimum eccentricity by solving $\dot{e_t^2} = 0$ in a post-Newtonian expansion to obtain the expression given in Eq.~(6) in Klein {\it et al.}~\cite{Klein:2018ybm}. This expression is used to generate residual eccentricity samples. We denote the distribution of residual eccentricities as $\{e_{\rm res}^{1000}(q, \boldsymbol{\chi}^{}_1, \boldsymbol{\chi}^{}_2)\}$, where we show the dependence on the binary's parameters here, for clarity. The details of the distributions of the initial binary parameters are given in Table~\ref{tab:BinarySimParams}. In our simulations of $50000$ binaries, all but $183$ binaries are successfully evolved to $a=10M$. The binaries whose evolution failed have initial eccentricities in the range $\sim 10^{-5}$--$10^{-4}$, all given by the residual eccentricity expression. The failures occur because in these cases the PN expression with leading order spin terms used for $e_{\rm res}^{1000}$ underestimates the minimum eccentricity obtainable  from $\dot{e_t^2}=0$ in our code.

\begin{figure}[t]
	\centering
	\includegraphics[width=.45\textwidth]{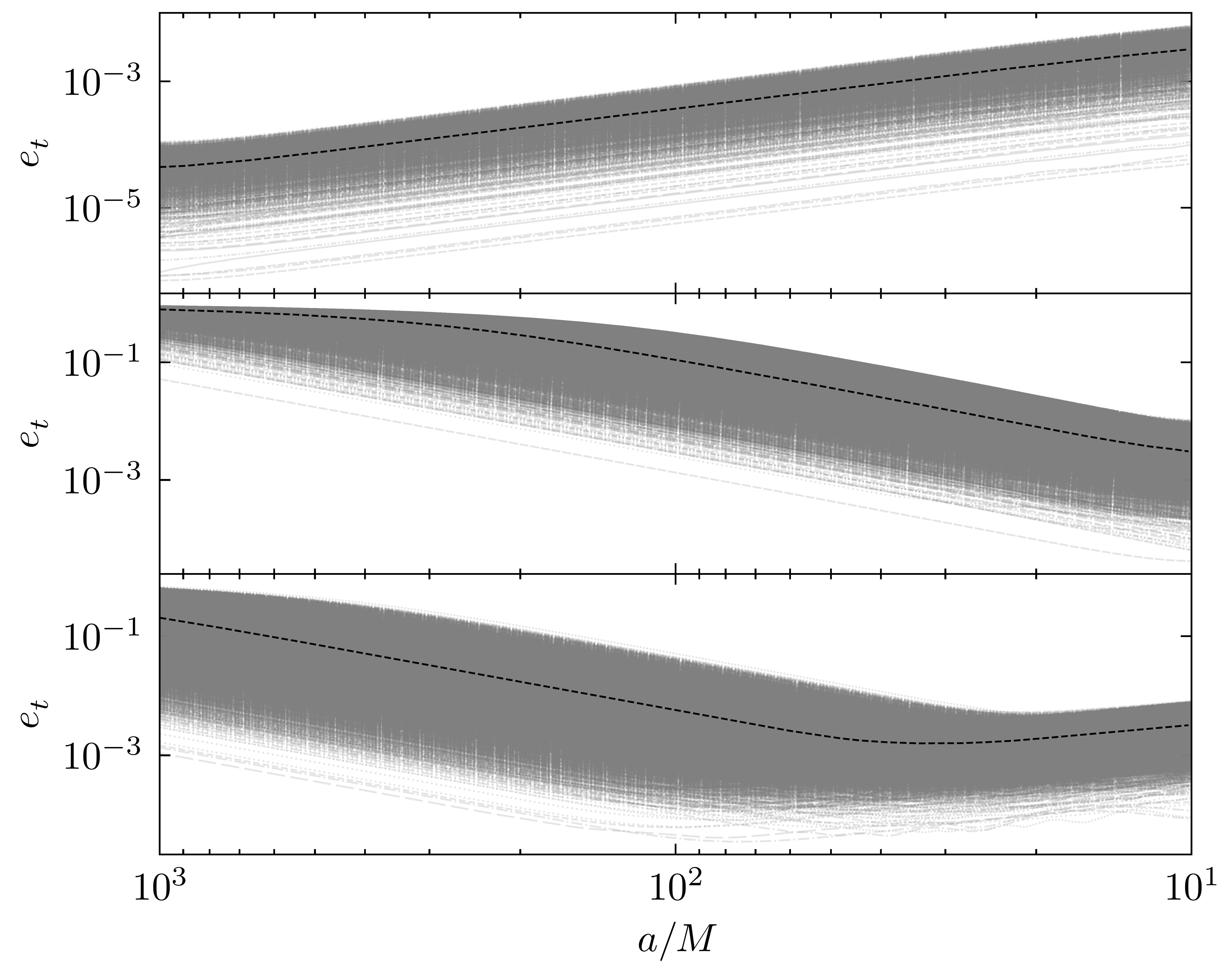}
	\caption{\label{fig:eccevolution}This figure illustrates the diversity in evolution of eccentricities in a population of precessing, eccentric binaries from an initial semimajor axis of $1000 M$ to a final semimajor axis of $10 M$. The panels show the eccentricity evolution of different binaries in the population, divided into three different possible scenarios: from top to bottom, monotonically increasing, monotonically decreasing, and nonmonotonic evolution, where the first and third scenarios are only possible due to the $2$PN spin-spin contributions to $\dot{e_t^2}$.
	The dotted line in each panel denotes the median value of eccentricity at a given separation in that subpopulation of binaries. }
\end{figure}

Similarly to Fig.~2 in~\cite{Phukon:2019gfh},  we present the results of our simulations for eccentricity evolution in three subplots in Fig.~\ref{fig:eccevolution}, dividing the binary population into three sub-groups based on their patterns of eccentricity evolution. In the top panel, eccentricities grow continuously during GW-driven inspiral of BBHs. This trend is observed for binaries with initial eccentricities similar to $e^{1000}_{\rm res}$.  As the spin-spin contributions to $\dot{e_t^2}$ are $\propto a^{-6}$, the residual eccentricites are also expected to grow as binary inspiral towards merger. It is noteworthy that the eccentricity growth observed in our numerical evolution for eccentric, precessing binaries in vacuum spacetime is purely induced by spin precession and differs from eccentricity growth under the influence of radiation reaction predicted in~\cite{Cutler:1994pb} for isolated (extreme mass-ratio) binaries and in~\cite{Cardoso:2020iji} for binaries interacting with surrounding matter.\footnote{Refs.~\cite{Lincoln:1990ji,Loutrel:2018ssg} have also predicted growth in eccentricity for isolated binaries in vacuum, however, this is a consequence of the use of the particular definition of eccentricity they use, as shown in~\cite{Ireland:2019tao,Will:2019lfe}.}.
The middle panel shows monotonic decay of eccentricity, the expectation due to radiation-reaction-driven backreaction on binaries in eccentric orbits~\cite{PhysRev.136.B1224}.
The bottom panel demonstrates the nonmonotonic evolution of eccentricities.   In this case, the eccentricities of the binaries initially decrease, reaching a minimum value, and subsequently increase again, possibly with oscillations (visible in the individual traces at the bottom of the panel), as were also found by~\cite{Klein:2010ti}. To make the trend visible, we only show binaries whose minimum eccentricity occurs for $a < 100M$. 
This diversity in eccentricity evolution trends in Fig.~\ref{fig:eccevolution} illustrates the complex behavior possible due to the interplay between eccentricity and spin precession.

\subsection{Spin Morphology Evolution}
\label{ssec:morph_evol}
Besides the impact of spins on eccentricity evolution, eccentricity itself can influence properties of spin dynamics during inspiral. As mentioned earlier, spin morphology is a property of precessing BBHs with both spins nonzero, defined by the conservative dynamics over a precessional cycle.  The spin parameter space for a binary at a given orbital separation can be divided into three non-overlapping regions or \emph{morphologies}: the circulating (C) morphology and two librating morphologies (L0 and L$\pi$). This classification is  based on the evolution of the  azimuthal angle between the two spins, $\phi^{}_{12}$, over a precession cycle. 

For a binary in the C morphology, $\phi^{}_{12}$ can evolve in the full range $[0, 2\pi]$  during binary's precessional evolution. In the librating morphologies L0 and  L$\pi$, $\phi^{}_{12}$ oscillates around 0 and $\pi$, respectively,  never reaching $\pi$ for  L0 and $0$ for  L$\pi$. The zero-amplitude limits of the librating morphologies are the spin-orbit resonances found in~\cite{Schnittman:2004vq}. During secular evolution under radiation reaction, a binary may transition from one  morphology to another. Ref.~\cite{Johnson-McDaniel:2023oea} showed that it should be possible to determine the spin morphologies of quasicircular precessing binaries from their GW signals with high statistical confidence in some cases.   

In general, for BBHs in circular orbits, most systems are in the C morphology at large orbital separations, and may transition to a librating morphology as they approach merger (and potentially even transition back to the C morphology---see~\cite{Gerosa:2015tea}, particularly their Fig.~14.  
Eccentricity, which enhances radiation reaction and influences the orientations of spin and orbital angular momentum vectors, is expected to alter the fraction of binaries in a given  spin morphology compared to that for binaries in circular orbits. Phukon {\it et al.}~\cite{Phukon:2019gfh} studied the impact of eccentricity on the fraction of binaries in different morphologies across semimajor axes in the range $a=1000M$ to $a=10M$, finding a negligible effect of eccentricity on the fraction of binaries residing in different morphologies and qualitatively similar behavior to the quasicircular case.  Fumagalli and Gerosa~\cite{Fumagalli:2023hde} studied the effect of eccentricity on the fraction of binaries in a given morphology when evolving binaries with small eccentricities close to merger back in time, finding that at orbital separations larger than $\sim 10^3 M$, eccentricity increases the number of binaries occupying  librating morphologies compared to the quasicircular case. This is expected, since the morphology depends on the magnitude of the orbital angular momentum and a nonzero eccentricity reduces the magnitude of the orbital angular momentum compared to the quasicircular case, thus making a larger semimajor axes with a nonzero eccentricity equivalent to a smaller semimajor axes in the quasicircular case. 

\begin{figure}[!t]
\includegraphics[width=.48\textwidth]{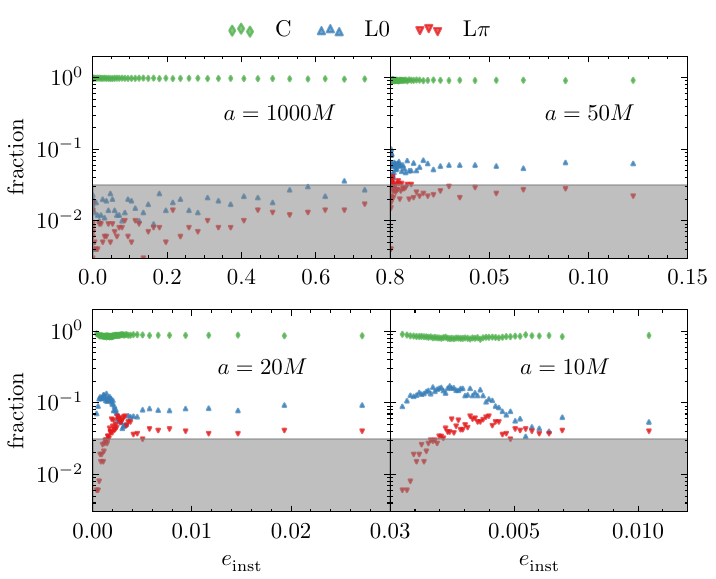}
\caption{\label{fig:morph_frac} Fraction of binaries in different morphologies as a function of instantaneous eccentricity $e_t^{}$ at a given semimajor axis. The eccentricities plotted are the upper edges  of each eccentricity bin in which the fraction of binaries in different morphologies are computed. The green, blue, and red points denote binaries in the C, L$0$, and L$\pi$ morphologies, respectively. The range of $e^{}_t$ values plotted on the horizontal axis decreases dramatically as the semimajor axis decreases.  The fractions of binaries in the three morphologies are computed in $50$ eccentricity bins with $\sim 1000$ binaries per bin.  The gray shaded regions indicate where the estimated fractions fall below the sampling error of $\sim 0.03$.}
\end{figure}

We use the simulated binaries from Sec.~\ref{ssec:ecc_evol} to investigate two aspects of the spin morphology of eccentric, precessing binaries. First, we examine how the morphologies occupied by binaries depend on their instantaneous eccentricity at various orbital separations.  For morphology computation, we apply the analytical method from~\cite{Johnson-McDaniel:2023oea}. This computation is based on~\cite{Gerosa:2015tea} and uses the effective potential method introduced in that paper to determine the values of $\phi^{}_{12}$ at the turning points of the binary's conservative precessional evolution, solving the resulting cubic equation analytically using trigonometric functions. The calculation in that paper is for quasicircular orbits, but the only change necessary to apply it to our eccentric case is to generalize the expression for the orbital angular momentum to include eccentricity. Here we use the Newtonian orbital angular momentum [Eq.~\eqref{eq:LNmag}] used in our precession equations. The other inputs are the binary's mass ratio and spin magnitudes and the spin angles at the point in the evolution at which we want to compute the morphology. Figure~\ref{fig:morph_frac} presents the fractions of binaries in different morphologies at four representative orbital separations: $a=1000M$, $50M$, $20M$, and $10M$. To estimate the fraction of binaries in the three morphologies at a given orbital separation, we divide the binaries into 50 bins of eccentricity with bin-edges at equal quantile intervals. This procedure leaves each eccentricity bin with approximately 1000 binaries. The fractions of binaries in different morphologies are computed for each eccentricity bin. In Fig.~\ref{fig:morph_frac}, we show the fraction of binaries in each morphology against the eccentricities corresponding to the upper bin edges. The grey-shaded regions mark areas where the estimated fractions are lower than the statistical uncertainty due to the finite number of binaries in a bin, estimated to be $ \sim 1/\sqrt{1000} \simeq 0.03$. 

At larger separations, the majority of the binaries exhibit the  C  morphology, although around $1\%$ of the binaries are in the librating L$0$ and L$\pi$ morphologies. The plots in the first row of Fig.~\ref{fig:morph_frac} show no strong dependence of morphology fractions on eccentricity at these separations. As binaries evolve toward merger, some of them transition from the C morphology to one of the librating morphologies.  The plots in the last row of Fig.~\ref{fig:morph_frac} illustrate the fractions of binaries in each morphology at semimajor axes of $a=20M$ and $10M$. At these separations, there is an order of magnitude increase in the fraction of binaries in the L$\pi$ morphology for $e^{}_\text{inst} \gtrsim 0.002$ compared to the smallest eccentricity bin. Additionally, the largest fraction of binaries in the L$0$ morphology is found for smaller eccentricities than the largest fraction in the L$\pi$ morphology.

Next, we investigate how eccentricity influences  orbital separations at which binaries initially in the  C morphology   transition to librating morphologies. A binary may transition back to the C morphology from a librating morphology (though this is not common---we find $160$ and $179$ cases with such multiple transitions to/from L$0$ and L$\pi$ in the set of $50000$ binaries considered in the previous subsection), but here we focus on the first transition from the C morphology to one of the librating morphologies. We refer to the semimajor axis at the transition as $a^{}_{\rm tr}$. To examine this, we identify  binaries that  are  in the  C morphology at $a=1000 M$ and undergo a morphology transition during the evolution. We classify these binaries  based  their morphology at the first transition (if there are multiple transitions), i.e., C$\rightarrow$L0 or C$\rightarrow$L$\pi$, and then identify binaries with the most symmetric masses from each subgroup. (Binaries with mass ratios close to unity are more likely to undergo a transition from the C morphology to a librating morphology~\cite{Gerosa:2015tea}. For this reason, we consider  the most symmetric binaries in each subgroup to increases the likelihood that a morphology transition will still occur even if one varies a parameter of the binary other than the mass ratio from its original value.) The parameters of the binary with the maximum mass ratio  undergoing the C$\rightarrow$L0 morphology transition  are $\{ q=0.999,\, e_0^{}=0.157,\, \chi_1^{}=0.196,\, \chi_2^{}=0.063,\, \theta_1^{}= 2.123~{\rm rad},\, \theta_2^{}=0.744~{\rm rad},\,\phi_1^{}=-0.704~{\rm rad}, \phi^{}_{2} = 5.476~{\rm rad} \}$. The  most symmetric binary in component masses undergoing C$\rightarrow$L$\pi$ transition has parameters $\{ q=0.998,\, e_0^{}=0.150,\, \chi_1^{}=0.931,\, \chi_2^{}=0.949,\, \theta_1^{}= 2.819~{\rm rad},\, \theta_2^{}=0.650~{\rm rad},\,\phi_1^{}=-0.755~{\rm rad},\, \phi^{}_{2} = 3.508~{\rm rad} \}$. The values of parameters are given at $a = 1000M$.

\begin{figure}[!t]
	\centering
	\includegraphics[width=.45\textwidth]{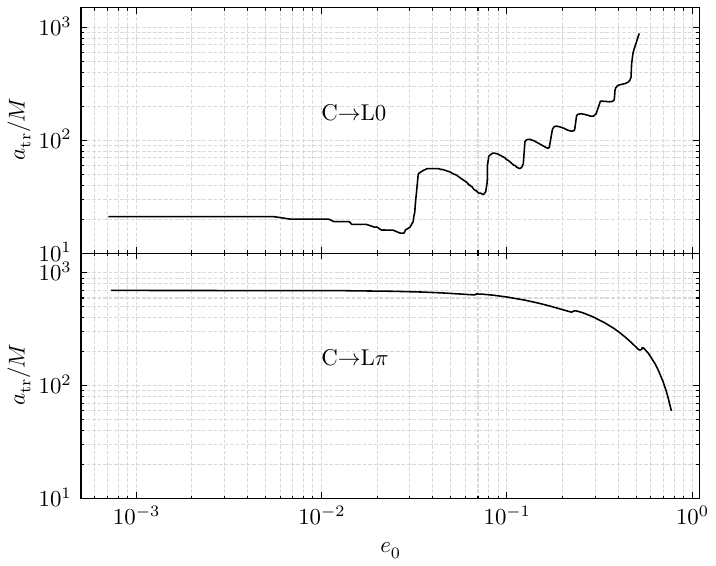}
	\caption{\label{fig:morph_trans_q} The dependence of the transition semimajor axis $a^{}_{\rm tr}$  on initial eccentricity, $e_0^{}$, starting at $a=1000M$.  The top panel shows a case that transitions to the L$0$ morphology, while the bottom panel shows a case that transitions to the L$\pi$ morphology. In both cases, the binary starts in the C morphology.  In the top panel, $257$ of the $300$ binaries undergo transition from the C morphology to the L0 morphology. The remaining binaries at eccentricities up to $0.97$ are in the L$0$ morphology at the initial separation. In the bottom panel, $298$ of the $300$ binaries transition from C to L$\pi$ morphology, with only $2$ binaries at large eccentricities starting in the L$0$ morphology.}
\end{figure}

We choose $300$ random initial eccentricities for each of those two binaries sampled from a uniform distribution in the range $e_0^{} \in [0, 0.97]$, while keeping the other parameters identical to their original configurations.  Most of these newly sampled  binaries are  in the  C morphology at the initial separation $a=1000M$ and we evolve them to $a=10M$ using the same setup as in Sec.~\ref{ssec:ecc_evol}. Out of these $300$ cases, $257$ binaries transition to the L$0$ morphology by $a = 10M$ for the first set of parameters, while for the second set of parameters, $298$ binaries transition to the L$\pi$ morphology by $a = 10M$. The binaries which do not transition from a circulating to a librating morphology during evolution start in the L$0$ morphology in both cases. 
In both cases, the binaries for which this occurs have large $e^{}_0$ values.  We plot $a^{}_{\rm tr}$ for the two sets of parameters versus initial eccentricity in Fig.~\ref{fig:morph_trans_q}. The top panel illustrates the transition to the L$0$ morphology (C$\rightarrow$L$0$), where it is observed that binaries with higher initial eccentricities undergo the transition at much larger orbital separations, even larger than $a = 1000M$ for $e_0$ slightly larger than $0.5$, which is where the plotted curve stops. One expects that the transition will occur at a larger semimajor axis as the eccentricity increases, since the morphology depends on the angular momentum, and this scales as $\sqrt{1 - e_t^2}$ (to Newtonian order, which is what is being used in this calculation). However, this scaling only gives an increase of $\sim 15\%$ for $e_t = 0.5$, while we find an increase of almost two orders of magnitude in $a^{}_{\rm tr}$ when increasing $e_0$. We also find a notable oscillatory dependence of $a^{}_{\rm tr}$ on $e^{}_0$. 
In contrast, the bottom panel depicts the transition to the L$\pi$ (C$\rightarrow$L$\pi$), which occurs at larger separation if the eccentricity is lower, and the binary instead transitions to the L$0$ morphology for eccentricities above $\sim 0.8$. 
For the cases in the bottom panel, we also find that for eccentricities above $\sim 0.673$, the binary transitions back to the C morphology before reaching $a = 10M$.

\section{Summary and Future Directions}
\label{sec:concl}

In this paper, we present a code 
for evolving precessing BBHs inspiraling on eccentric orbits using PN evolution equations from literature. The evolution of the orbital elements (such as orbital frequency, eccentricity, and periastron precession) follows the quasi-Keplerian parametrization~\cite{AIHPA_1985__43_1_107_0,AIHPA_1986__44_3_263_0} and is $3
$PN accurate in its point particle description (from~\cite{Arun:2007rg,Arun:2009mc}) with leading spin-orbit and spin-spin corrections from~\cite{Klein:2010ti} and quadrupole-monopole corrections (specialized to black holes) from~\cite{Klein:2018ybm}. The code has the option to choose the eccentricity-dependent enhancement functions in the fluxes either using the $O(e_t^4)$ expressions from Arun~\emph{et al.}~\cite{Arun:2009mc} or the high-accuracy super/hyperasymptotic expressions from Loutrel and Yunes~\cite{Loutrel:2016cdw}. The spin-induced precession of the orbital angular momentum and spin vectors is modeled using $2$PN accurate orbit-averaged precession equations from~\cite{Racine:2008qv}. The code can evolve binaries from large orbital separations to separations close to the merger or vice versa and uses the GNU Scientific Library implementation of the fourth-order adaptive Runge-Kutta-Fehlberg (4, 5) solver with additional Hermite interpolation, as implemented in LALSuite~\cite{LALSuite}. The code is implemented in the LALSimulation package of LALSuite and is currently under LVK review before it can be merged to master LALSuite. The code is written in C but can called using Python via the SWIGLAL-Python interface~\cite{Wette:2020air}.

In particular, the code either outputs full time series or just the values at a specified final frequency for $17$ quantities describing an eccentric, precessing BBH system. These are the average orbital frequency $\omega$, the eccentricity $e^{}_t$, the mean anomaly $l$, the mean orbital phase $\lambda$, the periastron advance parameter $k$, dimensionless  spin vector  components $\boldsymbol{\chi}^{}_{1,2}$, and  unit vectors of the orbital  angular momentum vector  $\mathbf{\hat{L}}$  and the  periastron direction $\boldsymbol{\hat{\cP}}$. To ensure physically meaningful results, the code applies multiple stopping conditions (detailed in Table~\ref{tab:stopcond}), terminating the evolution if unphysical behavior is detected, such as the binary's energy increasing with time. 

We have used this code to evolve eccentric, precessing BBHs and studied the effect of eccentricity on the binary dynamics. It is known that the $2$PN spin-spin term in the eccentricity evolution equation prevents the GW-driven circularization beyond a specific (mass- and spin-dependent) minimum value and even causes an increase in eccentricity during inspiral in some cases~\cite{Klein:2010ti}. In our simulations, we also studied the effect of including or not including the $2$PN spin-spin contributions to the evolution of $e^{}_{t}$. We found that if we start the evolution at $e^{}_{10\text{ Hz}}=0$, the $2$PN spin-spin contributions to the evolution of $e^{}_{t}$, the eccentricity at the end of the evolution when the stopping code is triggered can reach $e^{}_{t} \simeq 0.02$ in the cases we consider, while it remains $0$ when these contributions are turned off. 
Neglecting these $2$PN spin contributions to the evolution of $e^{}_{t}$ can lead to significant differences in other binary variables, notably $\psi^{}_1$, the angle between the in-plane component of spin 1 and the periastron line. 

On the other hand, we found negligible differences in binary parameter evolution if one uses higher order eccentricity terms in the hyperasymptotic enhancement functions. We further carried out a detailed study of how the choice of enhancement functions and their eccentricity truncation order influence the binary evolution. Our results indicate that hyperasymptotic enhancement functions~\cite{Loutrel:2016cdw} are necessary to obtain accurate dynamics at higher eccentricities, while the Arun~\emph{et al.}~\cite{Arun:2009mc} enhancement functions remain adequate at eccentricities $e^{}_{10 \text{Hz}}\lesssim 0.2$. We also used this code to replicate the three types of eccentricity evolution (monotonically increasing, monotonically deceasing, and decreasing and increasing after a minimum) for BBHs sub-populations considered in Phukon~\emph{et al.}~\cite{Phukon:2019gfh}. Finally, we illustrated how the binary's initial eccentricity affects the semimajor axis at which its spin morphology transitions from circulating to a librating morphology, and also that the fraction of binaries in a given spin morphology in a population depends on the binaries' instantaneous eccentricities.

The code developed in this paper can serve as a valuable tool for the astrophysics community in various ways.
In particular, the code will be useful in constraining astrophysical formation scenarios of BBHs based on their spin and eccentricity measurements in the sensitivity band of ground-based GW detectors. For example, using this code one can evolve the binary's spin tilts and eccentricity backward to the orbital separations at which the binary might have formed and compare the spin tilts and eccentricity at the time of formation with the predictions of various population synthesis and $N$-body simulations (e.g.,~\cite{DallAmico:2023neb}). Similarly, one can evolve the spins and eccentricity forward in time starting from population synthesis predictions and apply numerical relativity fits for eccentric, precessing BBHs (whenever those become available) to infer the properties of the BBHs' remnants and study the population of hierarchical mergers. However, a computationally  efficient  evolution of  binaries to/from large separations requires a more optimized approach, similar to that  exists for evolution of quasi-circular BBHs~\cite{Johnson-McDaniel:2021rvv,Gerosa:2023xsx}. We have now introduced a hybrid evolution code~\cite{Singh:2025sww} that combines the orbit-averaged evolution from this code with the precession-averaged evolution introduced in~\cite{Kesden:2014sla,Gerosa:2015tea} and generalized to the eccentric case in~\cite{Yu:2020iqj} (see also~\cite{Fumagalli:2023hde}) to efficiently evolve binaries backwards in time to compute their spin tilts at infinity.

Additionally, the code can be used in the development of eccentric, precessing waveform models for BBHs. Most directly, the binary dynamics computed by the code can then be used to compute time-domain GW polarizations, similar to the quasi-circular SpinTaylorT4 waveform (where~\cite{Buonanno:2002fy} describes the computation of the polarizations from the dynamics).  Such time-domain waveforms are valuable in assessing the performance of data analysis methods, particularly GW search algorithms, by performing injections of  waveforms in the data (see the use of SpinTaylorT2 waveforms in~\cite{Harry:2016ijz} for quasi-circular spin-precessing search development) or hybridizing with numerical relativity waveforms to create waveforms that cover the entire band of GW detectors (see, e.g.,~\cite{Chattaraj:2022tay,Sun:2024kmv} for hybridization of eccentric, nonspinning binaries and quasicircular, spin-precessing binaries, respectively). There already exist frequency-domain eccentric and precessing PN waveform models, notably the publicly available one from~\cite{Morras:2025nlp}, but these make various approximations to obtain the waveform in the frequency domain, so it would also be useful to have a publicly available time-domain waveform model.
The evolution of orbital angular momentum from the code will also be useful in approximating an eccentric, precessing waveform by twisting up a non-precessing eccentric waveform model (such as \cite{Nagar:2021xnh,Gamboa:2024hli,Planas:2025feq}), as is done (with a simple treatment of eccentricity plus precession) for the model in \cite{Nagar:2021xnh} in~\cite{Gamba:2024cvy}.

Finally, there are various improvements and checks of the code that we leave for future work. The angular momentum flux used in deriving the evolution equations sets the memory contribution to zero, following~\cite{Arun:2009mc}, while this contribution can potentially be quite significant for some formation scenarios, as discussed in Sec.~\ref{ssec:orb_evol_eqs}. Future work will include the memory contribution and study its effects on the evolution in astrophysically realistic scenarios in detail. Additionally, it will be important to include higher-order PN corrections for spin effects beyond the leading order terms currently present in the code. This will require extending the quasi-Keplerian description of binary orbits with misaligned spins beyond the current $2$PN order in spin terms. It will also be important to compare this with an evolution without orbit averaging, to assess the accuracy of this approximation, particularly for larger eccentricities. Here one could compare with a code that directly integrates the equations of motion, without using the quasi-Keplerian description, as in~\cite{Csizmadia:2012wy,Ireland:2019tao}; or using osculating orbits (as in~\cite{Loutrel:2018ydu,Fumagalli:2025rhc}, though these do not include the effects of spin); or by extending the closed-form solution to the $1.5$PN conservative dynamics given in~\cite{Samanta:2022yfe} to include higher-order terms and radiation reaction using perturbation theory.

\acknowledgments
We thank Lucy Thomas for carefully reading the manuscript and providing comments. We also thank Matthew Mould for useful comments.
Additionally, we thank Antoine Klein for confirming the typo we found and Maria Haney, Maria de Lluc Planas, Antoni Ramos Buades, and Lucy Thomas for useful discussions on the implementation of the code in LALSuite.
KSP acknowledges support from STFC grant ST/V005677/1. NKJ-M is supported by NSF grant AST-2205920. AS is supported by NSF grant PHY-2308887. AG is supported in part by NSF grants PHY-2308887 and AST-2205920.
The authors are grateful for computational resources provided by the LIGO Laboratory and supported by National Science Foundation Grants PHY-0757058 and PHY-0823459. 

This study used the software packages LALSuite~\cite{LALSuite}, matplotlib~\cite{Hunter:2007ouj}, numpy~\cite{Harris:2020xlr}, pandas~\cite{reback2020pandas,mckinney-proc-scipy-2010}, and scipy~\cite{2020SciPy-NMeth}.

This is LIGO document number P2500139.

\appendix

\section{Coefficients of orbital evolution equations}
\label{app:coeffs}

Here we give explicit expressions for the coefficients of the orbital evolution equations~\eqref{eq:evol}. The nonspinning coefficients come from~\cite{Arun:2009mc}, the spin-orbit terms from~\cite{Klein:2010ti}, and the spin-spin terms from~\cite{Klein:2018ybm}, where we have converted the expressions from the latter two references to the $\bar{x}$ expansion parameter we use. The PN expansion parameter $\bar{x}$ is defined in terms of eccentricity $e_t^{}$ and average orbital frequency $\omega$ as $ \bar {x} := \omega^{2/3}/(1 -
e_t^2)$.  We have also corrected a typo in the spin-spin contribution to $\dot{e_t^2}$ in Eq.~(A6b) in~\cite{Klein:2018ybm}, where in the fourth argument of $\sigma$, $111/4$ should be $111/2$. We found this while rederiving these expressions starting from the quadrupole-monopole terms in the energy and angular momentum fluxes given in~\cite{Gergely:2002fd} (added to the fluxes without these terms from~\cite{Klein:2010ti}) and got confirmation of this correction from A.~Klein~\cite{Klein_PC}. This correction was also found in~\cite{Morras:2025nlp} through comparison with the expressions in~\cite{Henry:2023tka}.

Specifically, we have
\begin{widetext}
\begin{subequations}
	{\allowdisplaybreaks
\begin{align}
\cO^{}_\text{N} &= \frac{96}{5} + \frac{292}{5}e_t^2 + \frac{37}{5}e_t^4,\\
\cO^{}_\text{1PN} &= -\frac{1486}{35} - \frac{264}{5}\eta + \left(\frac{2193}{7} - 570\eta\right)e_t^2 + \left(\frac{12217}{20} - \frac{5061}{10}\eta\right)e_t^4 + \left(\frac{11717}{280} - \frac{148}{5}\eta\right)e_t^6,\\
\nonumber
\cO^\text{hered, SO}_\text{1.5PN} &= \frac{384}{5}\pi(1 - e_t^2)^5\varphi(e^{}_t) + \frac{1}{30}\beta(-5424 - 27608 e_t^2 - 16694 e_t^4 - 585 e_t^6,\\
&\qquad\qquad \qquad \qquad \qquad \qquad  \;\;\, -3600 - 20928 e_t^2 - 14658 e_t^4 - 621 e_t^6),\\
\nonumber
\cO_\text{2PN}^\text{incl.\ SS} &= -\frac{11257}{945} + \frac{15677}{105}\eta + \frac{944}{15}\eta^2 + \left(-\frac{2960801}{945} - \frac{2781}{5}\eta + \frac{182387}{90}\eta^2\right)e_t^2 \\ \nonumber 
&\quad +  \left(-\frac{68647}{1260} - \frac{1150631}{140}\eta + \frac{396443}{72}\eta^2\right)e_t^4 + \left(\frac{925073}{336} - \frac{199939}{48}\eta + \frac{192943}{90}\eta^2\right)e_t^6 \\
\nonumber
&\quad  + \left(\frac{391457}{3360} - \frac{6037}{56}\eta + \frac{2923}{45}\eta^2\right)e_t^8\\
\nonumber
&\quad + \left[48 - \frac{96}{5}\eta + \left(2134 - \frac{4268}{5}\eta\right)e_t^2 + \left(2193 - \frac{4386}{5}\eta\right)e_t^4 + \left(\frac{175}{2} - 35\eta\right)e_t^6\right]\sqrt{1 - e_t^2}\\
\nonumber
&\quad + \frac{1}{320}\,\sigma(-15808 - 90400 e_t^2 - 63640 e_t^4 - 3084 e_t^6,\, 46144 + 259040 e_t^2 + 179880 e_t^4 + 8532 e_t^6,\\
\nonumber
&\qquad\qquad\;\;\, -131344 e_t^2 - 127888 e_t^4 - 7593 e_t^6, 896 + 8512 e_t^2 + 7728 e_t^4 + 504 e_t^6,\,  \\
&\qquad\qquad\;\;\, -128 - 1216 e_t^2 - 1104 e_t^4 - 72 e_t^6,\, 32 e_t^2 + 160 e_t^4 + 18 e_t^6),\\
\cO^\text{hered}_\text{2.5PN} &= -\pi(1 - e_t^2)^6\left[\frac{4159}{35}\psi^{}_\omega(e^{}_t) + \frac{2268}{5}\eta\,\zeta^{}_\omega(e^{}_t)\right],\\
\nonumber
\cO^\text{incl.\ hered}_\text{3PN} &= \frac{614389219}{148500} + \left(-\frac{57265081}{11340} + \frac{369}{2}\pi^2\right)\eta - \frac{16073}{140}\eta^2 - \frac{1121}{27}\eta^3\\
\nonumber
&\quad + \left[\frac{19769277811}{693000} + \left(\frac{66358561}{3240} + \frac{42571}{80}\pi^2\right)\eta - \frac{3161701}{840}\eta^2 - \frac{1287385}{324}\eta^3\right]e_t^2\\
 \nonumber
&\quad + \left[-\frac{3983966927}{8316000} + \left(\frac{6451690597}{90720} - \frac{12403}{64}\pi^2\right)\eta + \frac{34877019}{1120}\eta^2 - \frac{33769597}{1296}\eta^3\right]e_t^4\\
\nonumber
&\quad + \left[-\frac{4548320963}{5544000} + \left(-\frac{59823689}{4032} - \frac{242563}{640}\pi^2\right)\eta + \frac{411401857}{6720}\eta^2 - \frac{3200965}{108}\eta^3\right]e_t^6\\
\nonumber
&\quad + \left[\frac{19593451667}{2464000} + \left(-\frac{6614711}{480} - \frac{12177}{640}\pi^2\right)\eta + \frac{92762}{7}\eta^2 - \frac{982645}{162}\eta^3\right]e_t^8\\
\nonumber
&\quad + \left(\frac{33332681}{197120} - \frac{1874543}{10080}\eta + \frac{109733}{840}\eta^2 - \frac{8288}{81}\eta^3\right)e_t^{10}\\
\nonumber
&\quad + \biggl\{-\frac{1425319}{1125} + \left(\frac{9874}{105} - \frac{41}{10}\pi^2\right)\eta + \frac{632}{5}\eta^2 + \left[\frac{933454}{375} + \left(-\frac{2257181}{63} + \frac{45961}{240}\pi^2\right)\eta + \frac{125278}{15}\eta^2\right]e_t^2\\
\nonumber
&\quad + \left[\frac{840635951}{21000} + \left(-\frac{4927789}{60} + \frac{6191}{32}\pi^2\right)\eta + \frac{317273}{15}\eta^2\right]e_t^4\\
\nonumber
&\quad + \left[\frac{702667207}{31500} + \left(-\frac{6830419}{252} + \frac{287}{960}\pi^2\right)\eta + \frac{232177}{30}\eta^2\right]e_t^6 + \left(\frac{56403}{112} - \frac{427733}{840}\eta + \frac{4739}{30}\eta^2\right)e_t^8\biggr\}\sqrt{1 - e_t^2}\\
\nonumber
&\quad + \left(\frac{54784}{175} + \frac{465664}{105}e_t^2 + \frac{4426376}{525}e_t^4 + \frac{1498856}{525}e_t^6 + \frac{31779}{350}e_t^8\right)\ln\left(\frac{1 + \sqrt{1 - e_t^2}}{2}\bar{x}\right)\\
&\quad + \left(-\frac{3736352}{6125}\kappa(e^{}_t) + \left\{\frac{512}{5}\pi^2 - \frac{54784}{175}\left[\gamma + \ln(4\omega)\right]\right\}F(e^{}_t)\right)(1 - e_t^2)^{13/2},
\end{align}
	}
\end{subequations}
\begin{subequations}
	{\allowdisplaybreaks
\begin{align}
\cE^{}_\text{N} &= \left(\frac{304}{15} + \frac{121}{15}e_t^2\right)e_t^2,\\
\cE^{}_\text{1PN} &= \left[-\frac{939}{35} - \frac{4084}{45}\eta + \left(\frac{29917}{105} - \frac{7753}{30}\eta\right)e_t^2 + \left(\frac{13929}{280} - \frac{1664}{45}\eta\right)e_t^4\right]e_t^2,\\
 \cE^\text{hered, SO}_\text{1.5PN} &= \frac{394}{3}\pi e_t^2(1 - e_t^2)^4\varphi^{}_e(e^{}_t) - \frac{e_t^2}{90}\beta(19688 + 28256 e_t^2 + 2367 e_t^4,\, 13032 + 24270 e_t^2 + 2505 e_t^4),\\
 \nonumber
\cE_\text{2PN}^\text{incl.\ SS}\ &= \biggl\{-\frac{961973}{1890} + \frac{70967}{210}\eta + \frac{752}{5}\eta^2 + \left(-\frac{3180307}{2520} - \frac{1541059}{840}\eta + \frac{64433}{40}\eta^2\right)e_t^2,\\
\nonumber
&\quad + \left(\frac{23222071}{15120} - \frac{13402843}{5040}\eta + \frac{127411}{90}\eta^2\right)e_t^4 + \left(\frac{420727}{3360} - \frac{362071}{2520}\eta + \frac{821}{9}\eta^2\right)e_t^6\\
\nonumber
&\quad + \left[\frac{1336}{3} - \frac{2672}{15}\eta + \left(\frac{2321}{2} - \frac{2321}{5}\eta\right)e_t^2 + \left(\frac{565}{6} - \frac{113}{3}\eta\right)e_t^4\right]\sqrt{1 - e_t^2}\biggr\}e_t^2\\
\nonumber
&\quad + \frac{1}{960}\sigma(320 - 62752 e_t^2 - 101080 e_t^4 - 9420 e_t^6,\, -320 + 179936 e_t^2 + 287160 e_t^4 + 26820 e_t^6,\\
\nonumber
&\qquad\qquad\;\;\, -88432 e_t^2 - 161872 e_t^4 - 16521 e_t^6,\, -640 + 5440 e_t^2 + 11760 e_t^4 + 1080 e_t^6,\\
&\qquad\qquad\;\;\,\, 640 + 320 e_t^2 - 3120 e_t^4 - 360 e_t^6,\, 32 e_t^2 + 160 e_t^4 + 18 e_t^6),\\
\cE^\text{hered}_\text{2.5PN} &= -\pi e^2_t (1 - e_t^2)^5\left[\frac{55691}{210}\psi^{}_e(e^{}_t) + \frac{305072}{315}\eta\,\zeta^{}_e(e^{}_t)\right],\\
\nonumber
\cE^\text{incl.\ hered}_\text{3PN} &= \biggl\{\frac{54177075619}{6237000} + \left(\frac{7198067}{22680} + \frac{1283}{10}\pi^2\right)\eta - \frac{3000381}{2520}\eta^2 - \frac{61001}{486}\eta^3\\
\nonumber
&\quad + \left[\frac{6346360709}{891000} + \left(\frac{9569213}{360} + \frac{54001}{960}\pi^2\right)\eta + \frac{12478601}{15120}\eta^2 - \frac{86910509}{19440}\eta^3\right]e_t^2\\
\nonumber
&\quad + \left[-\frac{126288160777}{16632000} + \left(\frac{418129451}{181440} - \frac{254903}{1920}\pi^2\right)\eta + \frac{478808759}{20160}\eta^2 - \frac{2223241}{180}\eta^3\right]e_t^4\\
\nonumber
&\quad + \left[\frac{5845342193}{1232000} + \left(-\frac{98425673}{10080} - \frac{6519}{640}\pi^2\right)\eta + \frac{6538757}{630}\eta^2 - \frac{11792069}{2430}\eta^3\right]e_t^6\\
\nonumber
&\quad + \left(\frac{302322169}{1774080} - \frac{1921387}{10080}\eta + \frac{41179}{216}\eta^2 - \frac{193396}{1215}\eta^3\right)e_t^8\biggr\}e_t^2\\
\nonumber
&\quad + \biggl\{-\frac{22713049}{15750} + \left(-\frac{5526991}{945} + \frac{8323}{180}\pi^2\right)\eta + \frac{54332}{45}\eta^2\\
\nonumber
&\quad + \left[\frac{89395687}{7875} + \left(-\frac{38295557}{1260} + \frac{94177}{960}\pi^2\right)\eta + \frac{681989}{90}\eta^2\right]e_t^2\\
\nonumber
&\quad + \left[\frac{5321445613}{378000} + \left(-\frac{26478311}{1512} + \frac{3501}{2880}\pi^2\right)\eta + \frac{225106}{45}\eta^2\right]e_t^4 + \left(\frac{186961}{336} - \frac{289691}{504}\eta + \frac{3197}{18}\eta^2\right)e_t^6\biggr\} \\ \nonumber 
&\quad \times e_t^2\sqrt{1 - e_t^2} + \frac{730168}{23625}\frac{e_t^2}{1 + \sqrt{1 - e_t^2}}  + \left(\frac{1316528}{1575} + \frac{4762784}{1575}e_t^2 + \frac{2294294}{1575}e_t^4 + \frac{20437}{350}e_t^6\right)e_t^2\\ \nonumber 
&\quad \times \ln\left(\frac{1 + \sqrt{1 - e_t^2}}{2}\bar{x}\right) + \left(\left[\frac{89789209}{55125} - \frac{1398704}{1575}\ln 2 + \frac{156006}{175}\ln 3\right]\kappa^{}_e(e^{}_t)\right. \\
&\quad  + \left. \left\{-\frac{12304}{45} + \frac{1316528}{1575}\left[\gamma + \ln(4\omega)\right]\right\}F^{}_e(e^{}_t)\right)e_t^2 (1 - e_t^2)^{11/2},
\end{align}
	}
\end{subequations}
and
\begin{subequations}
	{\allowdisplaybreaks
\begin{align}
\cK^{}_\text{1PN} &= \frac{192}{5} + \frac{168}{5}e_t^2,\\
\cK^\text{SO}_\text{1.5PN} &= -\left( \frac{96}{5} + \frac{84e_t^2}{5}\right)\beta(4, 3),\\
\cK_\text{2PN}^\text{incl.\ SS} &= \frac{9124}{35} - \frac{1424}{5}\eta + \left(\frac{28512}{35} - \frac{3804}{5}\eta\right)e_t^2 + \left(\frac{10314}{35} - \frac{1017}{5}\eta\right)e_t^4 + \left(\frac{192}{5} + \frac{168e_t^2}{5}\right)\gamma^{}_1,\\
\cK^\text{hered}_\text{2.5PN} &= \frac{768}{5}\pi(1 - e_t^2)^{5}\varphi^{}_k(e^{}_t),\\
\nonumber
\cK^{}_\text{3PN} &= \frac{232082}{189} + \left(-\frac{131366}{21} + \frac{738}{5}\pi^2\right)\eta + \frac{13312}{15}\eta^2 + \left[\frac{2842199}{630} + \left(-\frac{1659934}{105} + \frac{1271}{10}\pi^2\right)\eta + \frac{29879}{5}\eta^2\right]e_t^2\\
\nonumber
&\quad + \left[\frac{1640713}{252} + \left(-\frac{1304524}{105} + \frac{5371}{320}\pi^2\right)\eta + \frac{54133}{10}\eta^2\right]e_t^4 + \left(\frac{1850407}{1680} - \frac{388799}{280}\eta + \frac{19573}{30}\eta^2\right)e_t^6\\
&\quad + \left[672 - \frac{1344}{5}\eta + \left(2436 - \frac{4872}{5}\eta\right)e_t^2 + \left(672 - \frac{1344}{5}\eta\right)e_t^4\right]\sqrt{1 - e_t^2}.
\end{align}
}
\end{subequations}
\end{widetext}
Here $\gamma$ is the Euler-Mascheroni gamma constant, we have set the gauge parameter $r^{}_0 = 1$, giving $x^{}_0 = 1$, and we have the following definitions of the functions used in the spin terms:
\begin{subequations}
{\allowdisplaybreaks
\begin{align}
\beta(a,b) &:= \left[a(\mathbf{S}^{}_1 + \mathbf{S}^{}_2) + b\left(q\mathbf{S}^{}_1 + \frac{\mathbf{S}^{}_2}{q}\right)\right]\cdot\mathbf{\hat{L}},\\
\nonumber
\sigma(a,b,c,d,e,f) &:= a\|\mathbf{s}\|^2 + b(\mathbf{s}\cdot\mathbf{\hat{L}})^2 + c\|\mathbf{s}\times\mathbf{\hat{L}}\|^2\cos 2\psi\\
\nonumber
&\quad\; + d(\|\mathbf{s}^{}_1\|^2 + \|\mathbf{s}^{}_2\|^2) \nonumber\\ 
&\quad\; + e\left[(\mathbf{s}^{}_1\cdot\mathbf{\hat{L}})^2 + (\mathbf{s}^{}_2\cdot\mathbf{\hat{L}})^2\right] \nonumber\\
&\quad\; + f\left(\|\mathbf{s}^{}_1\times\mathbf{\hat{L}}\|^2\cos 2\psi^{}_1 \right. \nonumber\\
&\left. \quad \quad \quad + \|\mathbf{s}^{}_2\times\mathbf{\hat{L}}\|^2\cos 2\psi^{}_2\right),
\label{eq:sigma}
\end{align}
	}
\end{subequations}
where $\mathbf{s}^{}_A := \mathbf{S}^{}_A/m^{}_A$ ($A\in\{1,2\}$) and $\mathbf{s} := \mathbf{s}^{}_1 + \mathbf{s}^{}_2$.

\subsection{Enhancement functions}
\label{app:enhancement}

Most of the enhancement functions appearing in the previous expressions cannot be computed exactly in closed form. The exceptions are
\begin{subequations}
\begin{align}
F(e) &= \left(1 + \frac{85}{6}e^2 + \frac{5171}{192}e^4 + \frac{1751}{192}e^6 + \frac{297}{1024}e^8\right)\nonumber \\
&\quad\times\left(1 - e^2\right)^{-13/2},\\
\tilde{F}(e) &= \left(1 + \frac{229}{32}e^2 + \frac{327}{64}e^4 + \frac{69}{256}e^6\right)\left(1 - e^2\right)^{-5},\\
F_e(e) &= \left(1 + \frac{2782}{769}e^2 + \frac{10721}{6152}e^4 + \frac{1719}{24608}e^6\right)\nonumber \\
&\quad\times\left(1 - e^2\right)^{-11/2}.
\end{align}
\end{subequations}
Additionally, the following enhancement functions can be expressed in terms of others:
\begin{subequations}
\begin{align}
\psi^{}_\omega(e) &= \frac{1344}{4159}\left[\frac{1 - 5e^2}{1 - e^2}\varphi(e) - \frac{4\,\tilde{\varphi}(e)}{(1 - e^2)^{3/2}}\right] \nonumber \\
&\quad+ \frac{8191}{4159}\psi(e),\\
\zeta^{}_\omega(e) &= \frac{583}{567}\zeta(e) - \frac{16}{567}\varphi(e),\\
\varphi^{}_e(e) &= \frac{192}{985}\frac{(1 - e^2)\varphi(e) - \sqrt{1 - e^2}\,\tilde{\varphi}(e)}{e^2},\\
\nonumber
\psi^{}_e(e) &= \frac{18816}{55691e^2}\left[\left(1 - \frac{11}{7}e^2\right)\varphi(e) - \left(1 - \frac{3}{7}e^2\right)\right. \nonumber \\
&\left. \quad \times\, \frac{\tilde{\varphi}(e)}{\sqrt{1 - e^2}}\right] \nonumber\\
&\quad + \frac{16382}{55691}\frac{(1 - e^2)\psi(e) - \sqrt{1 - e^2}\tilde{\psi}(e)}{e^2},\\
\nonumber
\zeta^{}_e(e) &= \frac{924}{19067e^2}\left[-(1 - e^2)\varphi(e) + \left(1 - \frac{5}{11}e^2\right) \right. \nonumber \\
&\left. \quad \times\, \frac{\tilde{\varphi}(e)}{\sqrt{1 - e^2}}\right] \nonumber\\
&\quad + \frac{12243}{76268}\frac{(1 - e^2)\,\zeta(e) - \sqrt{1 - e^2}\,\tilde{\zeta}(e)}{e^2},\\
\nonumber
\kappa^{}_e(e) &= \frac{(1 - e^2)\,\kappa(e) - \sqrt{1 - e^2}\,\tilde{\kappa}(e)}{e^2}\\
&\quad \times\, \left(\frac{769}{96} - \frac{3059665}{700566}\ln 2 + \frac{8190315}{1868176}\ln 3\right)^{-1},\\
\varphi^{}_k(e) &= \frac{\tilde{\varphi}(e)}{(1 - e^2)^{3/2}}.
\end{align}
\end{subequations}
These enhancement functions all reduce to $1$ for $e \to 0$, as do the others.
For the basic enhancement functions that cannot be expressed in closed form, we first give the $O(e_t^4)$ expressions from Arun~\emph{et al.}~\cite{Arun:2009mc} used in Phukon~\emph{et al.}~\cite{Phukon:2019gfh} and then give the equation references providing the high-accuracy superasymptotic and hyperasymptotic expressions from Loutrel and Yunes~\cite{Loutrel:2016cdw}. All three of these are available as options in the code, by setting \texttt{enhancementFunc} to $0$, $1$, and $2$, respectively. Specifically, the $O(e_t^4)$ expressions from Arun~\emph{et al.~\cite{Arun:2009mc}}  are
\begin{subequations}
\begin{align}
\varphi(e) &= 1+ \frac{2335}{192}e^2 + \frac{42955}{768}e^4 + O(e^6),\\
\psi(e) &= 1 - \frac{22988}{8191}e^2 - \frac{36508643}{524224}e^4 + O(e^6),\\
\zeta(e) &= 1 + \frac{1011565}{48972}e^2 + \frac{106573021}{783552}e^4 + O(e^6),\\
\nonumber
\kappa(e) &= 1 + \left(\frac{62}{3} - \frac{4613840}{350283}\ln 2 + \frac{24570945}{1868176}\ln 3\right)e^2\\
\nonumber
&\quad +  \left(\frac{9177}{64} + \frac{271636085}{1401132}\ln 2 - \frac{466847955}{7472704}\ln 3\right)e^4\\
&\quad + O(e^6),
\end{align}
\end{subequations}
\begin{subequations}
\begin{align}
\tilde{\varphi}(e) &= 1 + \frac{209}{32}e^2 + \frac{2415}{128}e^4 + O(e^6),\\
\tilde{\psi}(e) &= 1 - \frac{17416}{8191}e^2 - \frac{14199197}{524224}e^4 + O(e^6),\\
\tilde{\zeta}(e) &= 1 + \frac{102371}{8162}e^2 + \frac{14250725}{261184}e^4 + O(e^6),\\
\nonumber
\tilde{\kappa}(e) &= 1 + \left(\frac{389}{32} - \frac{2056005}{233522}\ln 2 + \frac{8190315}{934088}\ln 3\right)e^2\\
\nonumber
&\quad +  \left(\frac{3577}{64} + \frac{50149295}{467044}\ln 2 - \frac{155615985}{3736352}\ln 3\right)e^4\\
&\quad + O(e^6).
\end{align}
\end{subequations}
The superasymptotic and hyperasymptotic expressions are too lengthy to reproduce here, so we give the appropriate equation numbers in~\cite{Loutrel:2016cdw} in Table~\ref{tab:LY_eqs}, as well as the expressions (from~\cite{Arun:2007rg,Arun:2009mc}) for certain enhancement functions in terms of the others for which superasymptotic expressions are given in~\cite{Loutrel:2016cdw}:
\begin{subequations}
\label{eqs:extra_enh_func_relations}
\begin{align}
\psi(e) &= \frac{13696}{8191}\alpha(e) - \frac{16403}{24573}\beta(e) - \frac{112}{24573}\gamma(e),\\
\zeta(e) &= -\frac{1424}{4081}\theta(e) + \frac{16403}{12243}\beta(e) + \frac{16}{1749}\gamma(e),\\
\kappa(e) &= F(e) + \frac{59920}{116761}\chi(e),
\end{align}
\end{subequations}
where the same relations hold with tildes applied to all the enhancement functions.

\begin{table}
\caption{\label{tab:LY_eqs} The equations from Loutrel and Yunes~\cite{Loutrel:2016cdw} that give the superasymptotic expressions for the enhancement functions and the hyperasymptotic corrections to be added to these expressions to obtain the hyperasymptotic expressions. 
The cases where only a superasymptotic equation is given are those that give the individual pieces that contribute to $\psi(e)$, $\tilde{\psi}(e)$, $\zeta(e)$, and $\tilde{\zeta}(e)$ [cf.\ Eqs.~\eqref{eqs:extra_enh_func_relations}] while the hyperasymptotic corrections are given for the full quantities.}
\begin{tabular}{ccc}
\hline\hline
Quantity & Superasymptotic Eq.\ & Hyperasymptotic corr.\ Eq.\\
\hline
$\varphi(e)$ & (140) & (176)\\
$\alpha(e)$ & (169) & \\
$\beta(e)$ & (142) &\\
$\gamma(e)$ & (143) &\\
$\theta(e)$ & (152) &\\
$\chi(e)$ & (171) & (186)\\
$\psi(e)$ &  & (182)\\
$\zeta(e)$ &  & (184)\\
\hline
$\tilde{\varphi}(e)$ & (141) & (177)\\
$\tilde{\alpha}(e)$ & (170) & \\
$\tilde{\beta}(e)$ & (144) &\\
$\tilde{\gamma}(e)$ & (145) &\\
$\tilde{\theta}(e)$ & (153) &\\
$\tilde{\chi}(e)$ & (172) & (187)\\
$\tilde{\psi}(e)$ &  & (183)\\
$\tilde{\zeta}(e)$ &  & (185)\\
\hline\hline
\end{tabular}
\end{table}

\section{Corrected computation of $\dot{\Omega}$}
\label{app:dotOmega}

While Phukon~\emph{et al.}~\cite{Phukon:2019gfh} set $\dot{\Omega} = \|\boldsymbol{\omega}^{}_p\|$, citing K{\"o}nigsd{\"o}rffer and Gopakumar~\cite{Konigsdorffer:2005sc} (henceforth KG), this expression only holds in the equal-mass or single-spin cases (as is stated in KG). However, one can follow KG to derive the general expression, noting that [KG (4.24)]
\<
\mathbf{\hat{i}} = \frac{\mathbf{\hat{z}}\times\mathbf{\hat{L}}}{\|\mathbf{\hat{z}}\times\mathbf{\hat{L}}\|},
\?
where $\mathbf{\hat{i}}$ is the direction of the line of nodes, as illustrated in Fig.~1 of KG, and $\mathbf{\hat{z}} = \mathbf{\hat{J}}$, where $\mathbf{J} = \mathbf{L} + \mathbf{S}$ is the total angular momentum ($\mathbf{S} := \mathbf{S}^{}_1 + \mathbf{S}^{}_2$), as will become important later). One also has [KG (4.26a)]
\<
\mathbf{\hat{i}} = \mathbf{\hat{x}}\cos\Omega + \mathbf{\hat{y}}\sin\Omega.
\?
The latter expression gives $\dot{\Omega} = \mathbf{\hat{z}}\cdot(\mathbf{\hat{i}}\times\dot{\boldsymbol{\hat{\mathrm{i}}}})$, while the first expression gives
\<
\begin{split}
\mathbf{\hat{i}}\times\dot{\boldsymbol{\hat{\mathrm{i}}}} &= \frac{(\mathbf{\hat{z}}\times\mathbf{\hat{L}})\times(\mathbf{\hat{z}}\times\dot{\boldsymbol{\hat{\mathrm{L}}}})}{\|\mathbf{\hat{z}}\times\mathbf{\hat{L}}\|^2}\\
&= \frac{[\mathbf{\hat{z}}\cdot(\mathbf{\hat{L}}\times\dot{\boldsymbol{\hat{\mathrm{L}}}})]\mathbf{\hat{z}}}{\|\mathbf{\hat{z}}\times\mathbf{\hat{L}}\|^2}.
\end{split}
\?
We thus have
\<
\dot{\Omega} = \frac{\mathbf{\hat{z}}\cdot(\mathbf{\hat{L}}\times\dot{\boldsymbol{\hat{\mathrm{L}}}})}{\|\mathbf{\hat{z}}\times\mathbf{\hat{L}}\|^2}.
\?
We now note that $\dot{\boldsymbol{\hat{\mathrm{L}}}} = \boldsymbol{\omega}^{}_p\times\mathbf{\hat{L}}$, so $\mathbf{\hat{L}}\times\dot{\boldsymbol{\hat{\mathrm{L}}}} = \boldsymbol{\omega}_p - (\boldsymbol{\omega}^{}_p\cdot\mathbf{\hat{L}})\mathbf{\hat{L}}$ (i.e., $\boldsymbol{\omega}^{}_p$ projected perpendicular to the orbital angular momentum), thus giving
\<
\dot{\Omega} = \frac{\mathbf{\hat{z}}\cdot\boldsymbol{\omega}^{}_p - (\mathbf{\hat{z}}\cdot\mathbf{\hat{L}})(\boldsymbol{\omega}^{}_p\cdot\mathbf{\hat{L}})}{\|\mathbf{\hat{z}}\times\mathbf{\hat{L}}\|^2}.
\?
We now note that $\mathbf{\hat{z}} = \mathbf{\hat{J}}$ gives
\<
\mathbf{\hat{z}} = \frac{\mathbf{L} + \mathbf{S}}{\|\mathbf{L} + \mathbf{S}\|},
\?
so we have
\<
\mathbf{\hat{z}}\cdot\boldsymbol{\omega}^{}_p - (\mathbf{\hat{z}}\cdot\mathbf{\hat{L}})(\boldsymbol{\omega}^{}_p\cdot\mathbf{\hat{L}}) = \frac{\mathbf{S}\cdot[\boldsymbol{\omega}^{}_p - (\boldsymbol{\omega}^{}_p\cdot\mathbf{\hat{L}})\mathbf{\hat{L}}]}{\|\mathbf{L} + \mathbf{S}\|}.
\?
Thus,
\<\label{eq:dotOmega}
\dot{\Omega} = \frac{\mathbf{S}\cdot[\boldsymbol{\omega}^{}_p - (\boldsymbol{\omega}^{}_p\cdot\mathbf{\hat{L}})\mathbf{\hat{L}}]\|\mathbf{L} + \mathbf{S}\|}{\|\mathbf{S}\times\mathbf{\hat{L}}\|^2}.
\?

We can check that this reproduces KG (4.32), in the equal-mass or single-spin cases, noting that KG's $\Upsilon$ is our $\Omega$. In these cases [from KG (4.17a)] $\boldsymbol{\omega}^{}_p = \chi^{}_\text{so}\mathbf{J}/r^3 =: \alpha\mathbf{J}$ ($\chi^{}_\text{so}$ is defined in KG, but its definition is not needed here), so
\<
\begin{split}
\dot{\Omega}^{}_\text{KG} &= \frac{\alpha\mathbf{S}\cdot[\mathbf{S} - (\mathbf{S}\cdot\mathbf{\hat{L}})\mathbf{\hat{L}}]\|\mathbf{J}\|}{\|\mathbf{S}\times\mathbf{\hat{L}}\|^2}\\
&= \alpha\|\mathbf{J}\|,
\end{split}
\?
reproducing KG (4.32), where we have noted that $\|\mathbf{S}\times\mathbf{\hat{L}}\|^2 = \|\mathbf{S}\|^2 - (\mathbf{S}\cdot\mathbf{\hat{L}})^2$.

In more general cases, one has to worry about the fact that the denominator of Eq.~\eqref{eq:dotOmega} vanishes if $\mathbf{S}\,\|\,\mathbf{\hat{L}}$. One gets an $0/0$ indeterminate form in such cases. While in some cases the limit is finite (e.g., if one spin is aligned and one takes the limit as the other spin becomes aligned), in most cases one finds that the expression diverges, which is why we do not use this expression in our code. It is possible that one could use a different parameterization that avoids such divergences, but the method we use of directly evolving the vector giving the periastron direction is simpler.

\section{Expressions for energy and orbital angular momentum}
\label{app:EandJ}

Here we give the explicit expressions for the energy variable $\varepsilon = -2E/\eta$ and the magnitude of the orbital angular momentum $L$ up to $3$PN order that we use in some of the stopping conditions discussed in Sec.~\ref{ssec:stopping_cond}. (We set $M = 1$ in this appendix.) The time derivatives of these expressions are computed using the chain rule, i.e., $\dot{\cX} = (\partial\cX/\partial x)\dot{x} + (\partial\cX/\partial e_t^2)\dot{e_t^2}$, where $\cX$ is either $\varepsilon$ or $L$ and $x = \omega^{2/3}$, so $\dot{x} = (2/3)\omega^{-1/3}\dot{\omega}$.

The nonspinning contributions to $\varepsilon$ and $L$ come from~\cite{Arun:2009mc}, where we obtain $L$ from their expression for $j = \varepsilon L^2/\eta^2$ and note that the scalings with the symmetric mass ratio are already present in their $E$ and $J$ variables. (Since~\cite{Arun:2009mc} only consider nonspinning binaries, the total and orbital angular momenta are the same, which is why they use $J$ for what we call $L$.) The spin-orbit terms come from~\cite{Klein:2010ti} and the spin-spin terms from~\cite{Klein:2018ybm}. We write the expressions in terms of variables with the Newtonian terms scaled out, $\bar{\varepsilon} := \varepsilon/x$ and $\bar{L} := L/L^{}_\text{N}= x^{1/2}L/(\eta\sqrt{1 - e_t^2})$, giving
\begin{widetext}
\begin{subequations}
\begin{align}
\bar{\varepsilon} &= 1 + \cH^{}_\text{1PN} \bar{x} + \cH^\text{SO}_\text{1.5PN}\bar{x}^{3/2} + \cH^\text{incl.\ SS}_\text{2PN}\bar{x}^2 + \cH^{}_\text{3PN}\bar{x}^3,\\
\bar{L} &= 1 + \cL^{}_\text{1PN} \bar{x}  + \cL^\text{SO}_\text{1.5PN}\bar{x}^{3/2} + \cL^\text{incl.\ SS}_\text{2PN}\bar{x}^2 + \cL^{}_\text{3PN}\bar{x}^3,
\end{align}
\end{subequations}
with
\begin{subequations}
\begin{align}
\cH^{}_\text{1PN} &= -\frac{3}{4} - \frac{\eta}{12} + \left(-\frac{5}{4} + \frac{\eta}{12}\right)e_t^2,\\
\cH^\text{SO}_\text{1.5PN} &= \frac{2}{3}\beta(4,3),\\
\cH^\text{incl.\ SS}_\text{2PN} &= -\frac{67}{8} + \frac{35}{8}\eta - \frac{\eta^2}{24} + \left(-\frac{19}{4} + \frac{21}{4}\eta + \frac{\eta^2}{12}\right)e_t^2 + \left(\frac{5}{8} - \frac{5}{8}\eta - \frac{\eta^2}{24}\right)e_t^4 + (5 - 2\eta)(1 - e_t^2)^{3/2} - \gamma^{}_1,\\
\cH^{}_\text{3PN} &= -\frac{835}{64} + \left(\frac{18319}{192} - \frac{41}{16}\pi^2\right)\eta - \frac{169}{32}\eta^2 - \frac{35}{5184}\eta^3 + \left[-\frac{3703}{64} + \left(\frac{21235}{192} - \frac{41}{64}\pi^2\right)\eta - \frac{7733}{288}\eta^2 + \frac{35}{1728}\eta^3\right]e_t^2\nonumber\\
&\quad + \left(\frac{103}{64} - \frac{547}{192}\eta - \frac{1355}{288}\eta^2 - \frac{35}{1728}\eta^3\right)e_t^4 + \left(\frac{185}{192} + \frac{75}{64}\eta + \frac{25}{288}\eta^2 + \frac{35}{5184}\eta^3\right)e_t^6\nonumber\\
&\quad + \left\{\frac{5}{2} + \left(-\frac{641}{18} + \frac{41}{96}\pi^2\right)\eta + \frac{11}{3}\eta^2 + \left[-35 + \left(\frac{394}{9} - \frac{41}{96}\pi^2\right)\eta - \frac{\eta^2}{3}\right]e_t^2 + \left(\frac{5}{2} + \frac{23}{6}\eta - \frac{10}{3}\eta^2\right)e_t^4\right\} \nonumber \\
&\quad\times\sqrt{1 - e_t^2},
\end{align}
\end{subequations}
\begin{subequations}	
\begin{align}
\cL^{}_\text{1PN} &= \frac{3}{2} + \frac{\eta}{6} + \left(-\frac{3}{2} + \frac{5}{6}\eta\right)e_t^2,\\
\cL^\text{SO}_\text{1.5PN} &= -\beta\left(\frac{10}{3},\frac{5}{2} -\frac{ e_t^2}{2}\right),\\
\cL^\text{incl.\ SS}_\text{2PN} &= \frac{47}{8} - \frac{27}{8}\eta + \frac{\eta^2}{24} + \left(\frac{21}{4} - \frac{5}{6}\eta - \frac{3}{4}\eta^2\right)e_t^2 + \left(\frac{11}{8} - \frac{73}{24}\eta + \frac{5}{24}\eta^2\right)e_t^4 + \left[-\frac{5}{2} + \eta + (-5 + 2\eta)e_t^2\right]\nonumber \\
&\quad \times (1 - e_t^2)^{1/2}  + \gamma^{}_1,\\
\cL^{}_\text{3PN} &= \frac{155}{16} + \left(-\frac{3151}{48} + \frac{123}{64}\pi^2\right)\eta + \frac{25}{8}\eta^2 + \frac{7}{1296}\eta^3 + \left[\frac{1227}{16} + \left(-\frac{265}{2} + \frac{119}{128}\pi^2\right)\eta + \frac{787}{36}\eta^2 + \frac{95}{432}\eta^3\right]e_t^2\nonumber\\
&\quad + \left(\frac{169}{16} - \frac{115}{6}\eta + \frac{109}{9}\eta^2 + \frac{127}{432}\eta^3\right)e_t^4 + \left(-\frac{13}{48} + \frac{283}{48}\eta - \frac{71}{36}\eta^2 - \frac{25}{1296}\eta^3\right)e_t^6\nonumber\\
&\quad + \left\{-\frac{5}{4} + \left(\frac{641}{36} - \frac{41}{192}\pi^2\right)\eta - \frac{11}{6}\eta^2 + \left[-\frac{135}{4} + \left(\frac{2359}{36} - \frac{41}{96}\pi^2\right)\eta - \frac{34}{3}\eta^2\right]e_t^2 + \left(5 - \frac{14}{3}\eta - \frac{\eta^2}{3}\right)e_t^4\right\}\nonumber \\
&\quad \times\sqrt{1 - e_t^2},
\end{align}
\end{subequations}
\end{widetext}
where $\beta(a,b)$ and $\gamma^{}_1$ are defined in Eqs.~\eqref{eqs:betagamma}.

We also give explicit expressions for the time derivative of the scaled energy variable $\varepsilon$ (without the additional scaling by the Newtonian terms used to simplify the expression for the quantity itself above) and the orbital angular momentum variable $L$, where we define $\dot{L}_\text{sc} := \dot{L}/|\dot{L_\text{N}^{}}| = 2x^{3/2}\dot{L}/(\eta\sqrt{1 - e_t^2})$ to simplify the expression (since we only are interested in the sign of this quantity):
\begin{widetext}
\begin{subequations}
\begin{align}
\dot{\varepsilon} &=  \frac{\partial\varepsilon}{\partial x}\dot{x} + \frac{\partial\varepsilon}{\partial e_t^2}\dot{e_t^2}, \\
\frac{\partial\varepsilon}{\partial x} &= 1 + 2 \cH^{}_\text{1PN} \bar{x} + \frac{5}{2} \cH^\text{SO}_\text{1.5PN} \bar{x}^{3/2} + 3 \cH^\text{incl.\ SS}_\text{2PN} \bar{x}^2 + 4 \cH^{}_\text{3PN} \bar{x}^3  , \\
\frac{\partial\varepsilon}{\partial e_t^2} &= -2 \bar{x}^2 + \bar{x}^{5/2} \beta(4,3) - \frac{\bar{x}^3 }{2\sqrt{1-e_t^2}} \left\{-5 + 43 \sqrt{1-e_t^2} +  \left( 2-28 \sqrt{1-e_t^2}  \right)\eta + \left[10+7\sqrt{1-e_t^2}\, -  \right. \right. \nonumber \\
&\quad \left. \left. 4\left(1+2\sqrt{1-e_t^2}\right)\eta \right] e_t^2 + \left(-5 + 2\,\eta\right) e_t^4 \right\} - 2 \bar{x}^3 \gamma^{}_1 + \frac{\bar{x}^4 }{576 \sqrt{1-e_t^2}} \left\{ -16560 - 55872 \sqrt{1-e_t^2}\, + \right. \nonumber \\
& \quad  \left[ -26064 + 228576 \sqrt{1-e_t^2} - \left(-369 + 4797 \sqrt{1-e_t^2}\right)\pi^2\right] \eta + \left( 5088 - 24592 \sqrt{1-e_t^2} \right)\eta^2\, + \nonumber \\ 
&  \quad \left[-10800 - 64800 \sqrt{1-e_t^2} + \left(68304 + 124128 \sqrt{1-e^2_t}\right)\eta - 738\left( 1+\sqrt{1-e_t^2} \right)\pi^2\,\eta \, - \right. \nonumber \\ 
&  \quad \left.   \left( 9216 + 36352 \sqrt{1-e_t^2}\right)\eta^2  \right] e_t^2 + \left[ 28080 + 2592 \sqrt{1-e_t^2} + \left(-\, 41136 + 384 \sqrt{1-e_t^2} +369\pi^2\right) \eta\, -  \right. \nonumber  \\ 
& \quad  \left. \left. \left( -3168 + 2560 \sqrt{1-e_t^2} \right) \eta^2  \right] e_t^4 + \Big(-720 - 1104 \eta\,  + 960\eta^2 \Big)e_t^6 \right\},
\end{align}
\end{subequations}

\begin{subequations}
	\begin{align}
		\dot{L}_\text{sc} &=  \left(\frac{\partial L}{\partial x}\right)_\text{sc}\dot{x} + \left(\frac{\partial L}{\partial e_t^2}\right)_\text{sc}\dot{e_t^2}, \\
		\left(\frac{\partial L}{\partial x}\right)_\text{sc} &:= \frac{1}{|\dot{L_\text{N}^{}}|} \frac{\partial L}{\partial x} = - 1 + \cL^{}_\text{1PN} \bar{x} +  2\cL^\text{SO}_\text{1.5PN} \bar{x}^{3/2} + 3 \cL^\text{incl.\ SS}_\text{2PN} \bar{x}^2 + 5 \cL^{}_\text{3PN} \bar{x}^3, \\
		\left(\frac{\partial L}{\partial e_t^2}\right)_\text{sc} &:= \frac{1}{|\dot{L_\text{N}^{}}|} \frac{\partial L}{\partial e_t^2}= -\, \bar{x} + \frac{\bar{x}^2}{6}\Bigl[- 9 + 11 \eta + e_t^2 \left(9 - 5 \eta \right)\Bigr] + \bar{x}^{5/2} \Bigl[ 2 \cL^\text{SO}_\text{1.5PN}  + (1-e_t^2)\beta(0,1) \Bigr] + \frac{\bar{x}^3}{24}\Big\{ 675\, -  \nonumber \\
		& \quad \left. 360 \sqrt{1-e_t^2}  +\left( -283 + 144 \sqrt{1-e_t^2}\right)\eta -33\, \eta^2 + \Bigl( 258 -  312\, \eta +2 \eta^2  \Bigr) e_t^2 + \Bigl( - 33 +   73\, \eta - 5\, \eta^2 \Bigr)e_t^4 + 72\,\gamma_1^{} \right\} + \nonumber \\
		& \quad \frac{\bar{x}^4}{10368} \left\{  2092392 - 751680 \sqrt{1-e_t^2} - \left[ 6150600 - 2097216 \sqrt{1-e_t^2}  + \right. \right. \nonumber \\
		& \quad  \left. \left( -118908 + 17712 \sqrt{1-e_t^2} \right) \pi^2  \right] \eta -  \left( - 615312 + 311040 \sqrt{1-e_t^2} \right) \eta^2 + 4840\, \eta^3  +  \nonumber \\
		& \quad  \left[ 2823336 - 492480 \sqrt{1-e_t^2}\, + \left( -4916160 + 1165248 \sqrt{1-e_t^2}\right)\eta + \left( 28917 - 8856 \sqrt{1-e_t^2}\right)\pi^2\,\eta\, + \right. \nonumber \\
		& \quad  \left. \left( 1182240 - 248832 \sqrt{1 -e_t^2} \right) \eta^2 + 19032\, \eta^3 \right] e_t^2 + \Bigl( 92664 + 168048\, \eta + 2880\, \eta^2 + 1848\, \eta^3  \Bigr)\, e_t^4\, + \nonumber \\
		& \quad \Bigl( 2808 - 61128\, \eta + 20448\, \eta^2 + 200\, \eta^3  \Bigr)\,e_t^6  \biggr\}.
	\end{align}
\end{subequations}
\end{widetext}

\section{Example usage of code}
\label{app:example_usage}

The function \texttt{SimInspiralSpinTaylorPNEccentric-} \texttt{EvolveOrbit} evolves the binary, taking the following inputs (where the binary parameters are all in SI units): $\Delta T$, $m^{}_1$, $m^{}_2$, $f^{}_{\rm start}$, $f^{}_{\rm end}$, $\chi^{}_{1x}$, $\chi^{}_{1y}$, $\chi^{}_{1z}$, $\chi^{}_{2x}$, $\chi^{}_{2y}$, $\chi^{}_{2z}$, ${\hat L}^{}_x$, ${\hat L}^{}_y$, ${\hat L}^{}_z$, $e^{}_t$, $l^{}_0$, $\lambda^{}_0$, $\hat{\cP}^{}_x$, $\hat{\cP}^{}_y$, $\hat{\cP}^{}_z$, \texttt{spinO}, \texttt{EccOrder}, \texttt{phaseO}, \texttt{enhancementFunc}, \texttt{type}, and \texttt{LALparams}. Here, $\Delta T$ is the timestep for the interpolation (which is also used to initialize the adaptive timestepping in the integrator), while $m^{}_1$ and $m^{}_2$ are the binary's masses. The parameters $f^{}_{\rm start}$ and $f^{}_{\rm end}$ give the initial and final GW frequencies desired for the evolution; both $f^{}_{\rm start} < f^{}_{\rm end}$ (forward evolution) and $f^{}_{\rm start} > f^{}_{\rm end}$ (backward evolution) are possible. One can evolve forward as far as possible (until the evolution is stopped by one of the stopping conditions given in Table~\ref{tab:stopcond}) by setting $f^{}_{\rm end} = 0$. The spin and orbital angular momentum parameters are $\chi^{}_{1x}$, $\chi^{}_{1y}$, $\chi^{}_{1z}$, $\chi^{}_{2x}$, $\chi^{}_{2y}$, and $\chi^{}_{2z}$, the Cartesian components of the binary's dimensionless spins, and ${\hat L}^{}_x$, ${\hat L}^{}_y$, and ${\hat L}^{}_z$, the Cartesian components of the unit vector in the direction of the binary's (Newtonian) orbital angular momentum $\mathbf{L}$. The values of all these parameters are at $f^{}_\text{start}$, along with $e_t$, the binary's eccentricity, $l_0$ and $\lambda^{}_0$, the initial values of the mean anomaly and mean orbital phase, and $\hat{\cP}^{}_x$, $\hat{\cP}^{}_y$, $\hat{\cP}^{}_z$, the Cartesian components of the unit vector giving the location of the periastron (where $\boldsymbol{\hat{\cP}}$ must be orthogonal to $\mathbf{\hat{L}}$).

The remaining arguments determine various settings for the evolution. \texttt{spinO}${}\in\{0, 3, 4\}$ and \texttt{phaseO}${}\in \{0, 2, 3, 4, 5, 6\}$ denote twice the PN order of spin terms and nonspinning terms, respectively, in the evolution equations. While it is possible to pass \texttt{spinO} values of $1, 2$, these give the same results as a \texttt{spinO} value of $0$ since there are no spin terms at $0$PN, $0.5$PN, and $1$PN orders. Similarly, \texttt{phaseO} values of $0$ and $1$ give the same results as there are no $0.5$PN corrections to these equations. \texttt{enhancementFunc}${} \in\{0, 1, 2\}$ denotes the choice of enhancement function used in the evolution, where $0$ refers to expressions from Arun~\emph{et al.}~\cite{Arun:2009mc}, and $1$ and $2$ refer to the superasymptotic and hyperasymptotic expressions from Loutrel and Yunes~\cite{Loutrel:2016cdw}, respectively. \texttt{EccOrder}${}\in \{2, 4, 6, 8, 10, 12, 14, 16, 18, 20\}$ denotes the eccentricity order (highest power of $e_t$ used) in the hyperasymptotic enhancement function. To use the highest available PN order in the code, set all of \texttt{spinO}, \texttt{phaseO}, and \texttt{EccOrder} to $-1$. \texttt{type}${}\in\{0,1\}$ determines the type of output, as discussed below.

\texttt{LALparams} is a LAL dictionary containing additional settings for the code, and can be set to \texttt{None} if no such settings are needed. The three settings available are the ability to turn off the $2$PN spin-spin terms in the evolution of the eccentricity, so that orbits with an initial value of $e^{}_t = 0$ remain circular; to always output the results of the evolution even if it stops unexpectedly before reaching the desired final frequency (only available when just outputting the final value); and to set the initial value of $k$. These are set by inserting $0$ for the \texttt{EccEvol2PNSpinFlag} flag, $1$ for the \texttt{EccEvolAlwaysOutput} flag, and the desired initial value of $k$ for \texttt{InitialPeriastronPrecession}, respectively. See Table~\ref{tab:examples} for examples of all of these.

\texttt{SimInspiralSpinTaylorPNEccentricEvolve-} \texttt{Orbit} outputs the following quantities: $\omega$, $e^{}_t$, $l$, $\lambda$, $\chi^{}_{1x}$, $\chi^{}_{1y}$, $\chi^{}_{1z}$, $\chi^{}_{2x}$, $\chi^{}_{2y}$, $\chi^{}_{2z}$, ${\hat L}^{}_x$, ${\hat L}^{}_y$, ${\hat L}^{}_z$, $\hat{\cP}^{}_x$, $\hat{\cP}^{}_y$, $\hat{\cP}^{}_z$, and $k$. (Here $\omega$ is in total mass $= 1$ units, i.e., it is the same as $M\omega$.)
The type of output is determined by the \texttt{type} input, where $0$ corresponds to outputting the full time series and $1$ corresponds to outputting only the final value (at $f^{}_{\rm end}$). The final value of the full time series does not agree exactly with the output of just the final value, due to the specifics of how the evolution is stopped in the two cases, but the values approach each other as $\Delta T$ is decreased, as they must.

When a stopping condition besides the frequency bound is triggered, the code raises an error by default, except if one is evolving forward as far as possible, in which case the checks of the energy and angular momentum time derivatives, frequency second time derivative, and value of $\bar{x}$ are all allowed to stop the evolution with no error raised, in analogy with the similar behavior of the SpinTaylor evolution. If one wants to output the final value of the evolution up to the point at which an error is raised when evolving to a given final frequency, then it is necessary to set the \texttt{EccEvolAlwaysOutput} flag, as mentioned above.

We also tested the runtimes obtained when we evolved an ensemble of $10^{6}$ binaries where we sampled $m^{}_{1},m^{}_{2}$ randomly from $[5,200]M^{}_\odot$ and starting eccentricities from $[0,0.7]$. The spin magnitudes and spin angles were sampled randomly from their entire respective domains. The other settings like \texttt{spinO}, \texttt{phaseO}, \texttt{enhancementFunc} were all the same as in Sec.~\ref{ssec:ecc_evol}, except for the timestep for interpolation $\Delta T$ which was fixed at $1/8192$~s. We evolved the binaries forward in time from $10$~Hz using the full time-series option, as far as possible before triggering one of the stopping conditions. Most of the runtimes are below $1$~s with the largest being $5.2797$s, for the binary with $m^{}_1=170.938M_{\odot}, m^{}_2=20.656M_{\odot}$, $\chi^{}_1=0.557$, $\chi^{}_2=0.989$, $\theta^{}_{1}=0.824$~rad, $\theta^{}_{2}=2.255$~rad, $\phi^{}_{1}=2.059$~rad, $\phi^{}_{2}=2.159$~rad, and $e_{10\text{ Hz}}=0.2888$. These tests were performed on a heterogeneous computing pool comprising a range of CPUs, including Intel's Skylake-X (2017) and more recent models, as well as AMD's Zen~4 (2022) and later generations, with base clock speeds ranging from 2.2~GHz to 4.7~GHz.

In Table~\ref{tab:examples}, we give examples of the use of the code to evolve a binary, illustrating the use of the full time-series and only final value options, as well as the option to output the final value right before the code stops and giving the setting used to evolve forward as far as possible. We also show how to set up \texttt{LALparams} to turn off the $2$PN spin terms in eccentricity evolution equation and how to provide a desirable initial value of $k$.  We do this for a binary with masses $m^{}_1 = 50M_\odot$, $m^{}_2 = 45M_\odot$, dimensionless spin magnitudes $\chi^{}_1 = 0.8$, $\chi^{}_2 = 0.6$, and spin angles $\theta^{}_1 = 1.3$~rad, $\theta^{}_2 = 0.4$~rad, $\phi^{}_1 = 2.1$~rad, $\phi^{}_2 = 1.3$~rad, and eccentricity $e^{}_{20\text{ Hz}} = 0.3$ at $f^{}_\text{start} = 20$~Hz. The stopping code of $1032$ means that the code stopped because the magnitude of the orbital angular momentum started increasing (see Table~\ref{tab:stopcond}). The use of \texttt{s1*}, \texttt{s2*} as the names for the dimensionless spin components is for uniformity with the quasicircular SpinTaylor code~\cite{SpinTaylorEvolveOrbit}; we keep the same notation for the output in our example for brevity. 

\clearpage

\begin{widetext}

\captionof{table}{\label{tab:examples}
	Examples of using the code introduced in this paper~\cite{amiteshgit} to evolve binaries in an interactive Python session, using the highest PN order expressions available.
}
\lstset{language=Python}
\begin{lstlisting}
# Setup and initial evolution
>>> import numpy as np
>>> import lalsimulation as lalsim
>>> import lal
>>> m1, m2 = 50., 45. # solar masses
>>> chi1, chi2 = 0.8, 0.6
>>> tilt1, tilt2, phi1, phi2 = 1.3, 0.4, 2.1, 1.3 # rad
>>> fstart, fend = 20., 100. # Hz
# If one wants the evolution as far forward as possible, use fend = 0
>>> e0 = 0.3        # starting eccentricity at 20 Hz
>>> dictparams = {'deltaT': 1e-5, 'm1_SI': m1*lal.MSUN_SI, 'm2_SI': m2*lal.MSUN_SI, 'fStart': fstart,
	'fEnd': fend, 's1x': chi1*np.sin(tilt1)*np.cos(phi1), 's1y': chi1*np.sin(tilt1)*np.sin(phi1),
	's1z': chi1*np.cos(tilt1), 's2x': chi2*np.sin(tilt2)*np.cos(phi2),
	's2y': chi2*np.sin(tilt2)*np.sin(phi2), 's2z': chi2*np.cos(tilt2), 'lnhatx': 0., 'lnhaty': 0.,
	'lnhatz': 1., 'eccentricity': e0, 'l0': 0, 'lamb0': 0, 'phatx':1, 'phaty':0, 'phatz':0, 'spinO': -1,
	'EccOrder': -1, 'phaseO': -1, 'enhancementFunc': 2, 'LALparams': None}

# Evolve forward from fstart = 20 Hz to fend = 100 Hz
>>> dictparams['type'] = 0   # to get full timeseries output
>>> omega, et, l, lamb, s1x, s1y, s1z, s2x, s2y, s2z, LNhatx, LNhaty, LNhatz, Phatx, Phaty, Phatz, k\
     = lalsim.SimInspiralSpinTaylorPNEccentricEvolveOrbit(**dictparams)
>>> et.data.data
array([0.3       , 0.29999401, 0.29998803, ..., 0.02464491, 0.02455651, 0.02446785])
\end{lstlisting}

\begin{lstlisting}
# Evolve forward past the point where the code stops
# Here we have run "export LAL_DEBUG_LEVEL=7" on the command line before starting the Python session to get useful messages printed
>>> LALdict = lal.CreateDict()
# Output regardless of where the code stops
>>> lalsim.SimInspiralWaveformParamsInsertEccEvolAlwaysOutput(LALdict,1) 
# Evolve to 150 Hz, evaluating only the final value of the output (since type=1)
>>> dictparams['type'], dictparams['fEnd'], dictparams['LALparams'] = 1, 150, LALdict
>>> omega, et, l, lamb, s1x, s1y, s1z, s2x, s2y, s2z, LNhatx, LNhaty, LNhatz, Phatx, Phaty, Phatz, k\
     = lalsim.SimInspiralSpinTaylorPNEccentricEvolveOrbit(**dictparams)
XLAL Warning - XLALSimInspiralSpinTaylorPNEccentricEvolveOrbit: integration terminated with code 1032.
The final GW frequency reached was 114.651, while the desired final frequency is 150.
Evolution parameters were m1 = 5.000000e+01, m2 = 4.500000e+01,
s1 = (-3.891589e-01,6.654020e-01,2.139991e-01), s2 = (6.250137e-02,2.251363e-01,5.526366e-01),
ecc = 3.000000e-01
>>> et.data.data
array([0.01816366])
\end{lstlisting}

\begin{lstlisting}
# Switch off the 2PN spin-spin contributions to the evolution of eccentricity
>>> lalsim.SimInspiralWaveformParamsInsertEccEvol2PNSpinFlag(LALdict, 0)
# Evolve to 100 Hz, evaluating only the final value of the output
>>> dictparams['fEnd'], dictparams['LALparams'] = 100, LALdict
>>> omega, et, l, lamb, s1x, s1y, s1z, s2x, s2y, s2z, LNhatx, LNhaty, LNhatz, Phatx, Phaty, Phatz, k\
     = lalsim.SimInspiralSpinTaylorPNEccentricEvolveOrbit(**dictparams)
XLAL Info - XLALSimInspiralSpinTaylorPNEccentricEvolveOrbit: integration terminated with code 1029.
The final GW frequency reached was 99.9244
>>> et.data.data
array([0.02377233])
\end{lstlisting}

\begin{lstlisting}
# Insert a user-defined value for the initial value of k (the spin-spin contributions to eccentricity are still turned off)
# First compute the default value of k, for comparison [the default value in this case is 0.464..., as given by lalsim.SimInspiralSpinTaylorEccentricComputePeriastronPrecession()]
>>> lalsim.SimInspiralWaveformParamsInsertInitialPeriastronPrecession(LALdict, 2)
>>> omega, et, l, lamb, s1x, s1y, s1z, s2x, s2y, s2z, LNhatx, LNhaty, LNhatz, Phatx, Phaty, Phatz, k\
     = lalsim.SimInspiralSpinTaylorPNEccentricEvolveOrbit(**dictparams)
XLAL Info - XLALSimInspiralSpinTaylorPNEccentricEvolveOrbit: integration terminated with code 1029.
The final GW frequency reached was 99.9439
>>> et.data.data, k.data.data
array([0.02375917]), array([3.03367557])
\end{lstlisting}

\vspace{1cm}

\end{widetext}

\bibliography{ecc_refs}

\end{document}